\documentclass[11pt,a4paper]{scrartcl}

\usepackage[utf8]{inputenc}
\usepackage[english]{babel}           
\usepackage{amsfonts, amsmath, amsthm, amssymb, dsfont}

\usepackage{newtxmath}

\usepackage{graphicx}
\usepackage{epstopdf}
\usepackage{color}
\usepackage{hyperref}

\usepackage{lineno}

\usepackage{natbib}

\newcommand{\s}{\footnotesize}  %controll the font size of figure labels
\newcommand{\Real}{\Re}
\DeclareMathAlphabet\mathsfbi{T1}{phv}{b}{it}

\title{Waves in the gas centrifuge: asymptotic theory and similarities with the atmosphere}
\author{Marie Rodal and Mark Schlutow\footnote{mark.schlutow@fu-berlin.de}\\ 
\s Institut f\"ur Mathematik, Freie Universit\"at Berlin, Germany}
\date{}

\begin{document}

	\maketitle

	\begin{abstract}
		We study the stratified gas in a rapidly rotating centrifuge as a model for the Earth's atmosphere.
Based on methods of perturbation theory, 
it is shown that in certain regimes, internal waves in the gas centrifuge 
have the same dispersion relation to leading order as their atmospheric siblings.
Assuming an air filled centrifuge with a radius of circa 50\,cm, the optimal rotational frequency for realistic atmosphere-like waves
%that encounter altitudinal amplification, 
is around 10\,000 rounds per minute.
Using gases of lower heat capacities at constant pressure, like xenon, 
the rotational frequencies can be even halved to obtain the same results.
Similar to the atmosphere, it is feasible in the gas centrifuge to generate a clear scale separation of wave frequencies 
and therefore phase speeds between acoustic waves and internal waves.
In addition to the centrifugal force, the Coriolis force acts in the same plane. However, its influence on axially homogeneous internal waves
appears only as a higher-order correction.
We conclude that the gas centrifuge provides an unprecedented opportunity to investigate atmospheric internal waves experimentally 
with a compressible working fluid.
	\end{abstract}

	\hrule
	
	\hspace{8mm}

	``What was once thought can never be unthought.''
	\begin{flushright} 
		-- Friedrich Dürrenmatt (1962)
	\end{flushright}
	%\hspace{3mm}

	\section{Introduction}

%The importance of GW for the atmospheric circulation
Whenever gravity acts on a fluid of inhomogeneous density such that the fluid stably stratifies,
gravity waves may be excited. 
Indeed, Earth's atmosphere is for the most part stably stratified 
and gravity waves are an omnipresent oscillation mode.
They are usually excited in the troposphere 
wherefrom they propagate into the higher layers 
and interact with the mean flow by various processes \citep{Fritts2003,Alexander2010}.
Gravity waves transport energy away from their source 
and redistribute it elsewhere when becoming unstable which leads to wave breaking, 
ultimately turbulence and mixing \citep{Becker2012,Schlutow2014a}.
But most significant is the wave drag that acts as a body force on the mean flow
causing an acceleration of the mean-flow wind.
It is these interactions that make gravity waves so important 
for the atmospheric circulation.
Despite their importance for weather and climate forecast \citep{Kim2003,Orr2010,Lott2013,Kim2020},
considerable gaps in our understanding of the dynamics of gravity waves persist.

% About the theory of gravity waves and debate 
In particular the question of wave stability---when exactly do waves become unstable---remains 
for the most part unsettled. 
It adds to the complications that instability mechanisms are inherently nonlinear.
In the literature of theoretical fluid dynamics
several of those mechanisms have been proposed. 
For instance, waves destabilize due to static instabilities 
when they push denser air mass on top of lighter fluid
which is also associated with overturning and leads to breaking.

Waves may also become unstable due to the wind shear that they induce. 
The mechanism is similar to Kelvin-Helmholtz instabilities \citep{Fritts1985}.
Moreover, perturbative modes that form a triad with a gravity wave satisfying certain conditions 
blow up by the Parametric Subharmonic Instability (PSI); 
subharmonic means that the frequency of the perturbation 
is comparatively small with respect to the base wave 
and that they differ only by rational factors \citep{Mied1976a}.

In recent years, modulational instabilities have also gained some attention in the community.
This type of instability manifests itself in the modulation properties of the wave.
The evolution of waves and in particular of wave packets can be effectively described 
by the spatio-temporal evolution of their amplitude and phase which is governed by modulation equations. 
In certain conditions the amplitude or rather wave envelope %for instance 
may blow up due to modulational instabilities.
It was shown in the pioneering work of \cite{Grimshaw1972} that stationary plane gravity waves of large amplitudes
destabilize due to modulation. 
\cite{Schlutow2018} extended these ideas to classes of traveling wave solutions. 
Modulational instabilities of a primary wave may even cause the excitation of new, 
secondary waves which was proven to be theoretically possible in \cite{Schlutow2019}.
A generalized modulation theory studying the stability of almost unconstrained stationary gravity waves 
resembling mountain lee waves was presented in \cite{Schlutow2020a}.

It holds for all instability mechanisms that 
the onset of instability or the instability growth rate or both depend sensitively on the wave's amplitude.
In fact, the interplay of onset and growth rate are of utmost importance for the fate of the wave.
In the atmosphere the amplitude of gravity waves 
is for the bigger part controlled by a process that we call altitudinal amplification.
As gravity waves propagate upwards they encounter an exponentially decreasing background density.
Due to energy conservation, the amplitude must in turn increase also exponentially.
This process is a direct consequence of the compressibility of air.
The crux of the stability problem for atmospheric gravity waves is
the complicated dynamics due to altitudinal amplification 
and the nonlinear processes that emerge from it.

In order to verify or falsify the theoretical predictions, experiments and observations are essential.
Experiments by means of numerical simulations of the Navier-Stokes equations 
were performed by \cite{Sutherland2006a,Achatz2007,Dong2020} and many more.
Two basic ideas were exploited. On the one hand, a small domain with periodic boundaries of the size of exactly one wavelength of a monochromatic wave was used which allowed Direct Numerical Simulations (DNS).
This approach excludes all perturbations of larger size than the domain.
And on the other hand, large domains were utilized
allowing for large-scale dynamics
which however has the drawback that DNS become too expensive 
and remedies like Large Eddy Simulations (LES) need to be applied.

% Little eoverview about field obsevation and the problem of reproducability
The main driver for our understanding of gravity waves are certainly field measurements.
Field campaigns like DEEPWAVE \citep{Fritts2016,Fritts2019}
resulted in new insights into nonlinear wave dynamics.
Sophisticated devices like the HALO research aircraft \citep{Bramberger2020} 
or sounding satellites like HIRDLS \citep{Ern2018} provide a constant stream of new findings on gravity waves. 
Admittedly, those campaigns and measurement devices are expensive but most importantly they do not provide repeatable outcomes as the atmosphere is a chaotic system. 
Therefore, repeatable laboratory experiments are an indispensable, complementary 
and also usually much less expensive tool to corroborate theoretical predictions. 

% Laboratory experiments with GW using water
Gravity waves have been observed and studied in laboratories.
\cite{Rodda2020} investigated the excitation of gravity waves by baroclinic jets in a rotating annulus.
Carefully controlled gravity waves were measured with the Schlieren method by \cite{Sutherland2014}.
In the comprehensive study by \cite{Sutherland2013b} it is reported how PSI have been excited and observed in a laboratory.
The stability of gravity wave beams was explored experimentally by \cite{Bordes2012,Dauxois2018}.

% Limitations due to incompressibility and the need to study compressible waves
To our knowledge, all laboratory experiments on gravity waves were performed with water 
as working fluid where the stratification is obtained by controlling the temperature and salinity, accordingly.
Due to flow similarities, 
most of the features observed in the water tanks are equally valid for the atmosphere. 
However, one particular property of air cannot be emulated by water: compressibility. 
%This statement is especially true for gravity waves and their stability as we have argued above.
As we have argued above, it is in particular those processes caused by compressibility 
that determine the stability and hence the fate of atmospheric gravity waves.

\begin{figure}
	\centering
	\includegraphics[scale=0.8]{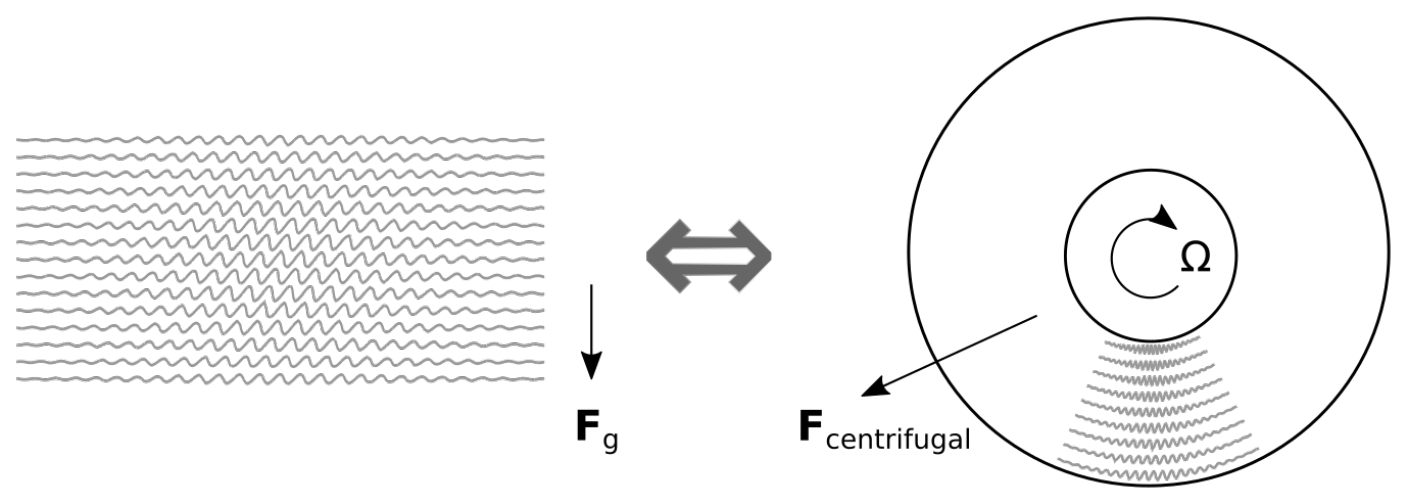}
	\caption{Sketch of the underlying idea. The gravitational pull from Earth (left) 
		is substituted by the centrifugal force in
		a centrifuge (right) which equally causes stratification supporting internal waves.}
	\label{fig:idea}
\end{figure}
%
% General idea, results of this paper and organization of the content
In this paper, we want to propose an experimental device to study compressible gravity waves.
The challenge of putting the atmosphere into a laboratory is posed by the density scale height being a measure for stratification. It is of the order of magnitude of 10\,km which is very weak. To obtain realistic waves in an experiment, a tank of the size of the scale height would be necessary which is obviously unfeasible. 
A feasible substitute could be a gas centrifuge. Here, the acceleration due to gravity is replaced by the centrifugal force. This idea is sketched in Figure~\ref{fig:idea}.

The goal of this paper is to study the waves in gas centrifuges theoretically and 
point out the similarities to atmospheric gravity waves.
In particular, we will show that the gas centrifuge supports the same
features caused by compressibility like the altitudinal amplification.
%We will provide explicit dispersion and polarization relations.
The outcomes of this study shall be the basis for an unprecedented experiment.

Bogovalov and collaborators \citep{Bogovalov2015,Bogovalov2019,Bogovalov2020} 
have made substantial contributions to the literature on the theory of waves 
in gas centrifuges at high rotational velocities. 
In \cite{Bogovalov2015} they explore the behavior of inviscid axisymmetric waves
that propagate in the radial and axial direction. 
They derive an analytical solution in terms of Whittaker functions, and explore their dispersive properties. 
%In addition, an expression is derived for the corrections to the perturbation pressure due to centrifugal field.
%The waves thusly described are equivalent to evanesecent waves in the atmosphere, which propagate exclusively in the horizontal direction, and whose amplitude increases with altitude; as the restoring force, i.e. the centrifugal force, acts in the radial direction, the axis are in the case of the centrifuge flipped.
In addition, \cite{Bogovalov2019} advanced the theory by also incorporating dissipation into the model. 
We are building on the work done by Bogovalov and collaborators 
by first removing the restriction that the waves be axisymmetric, 
and then using asymptotic perturbation methods to obtain
closed form formulas for the dispersion and polarization relations.
Moreover, we will include the effects of gravity to confirm 
that its influence remains negligible. 
%
%Waves in rapidly rotating gas centrifuges were explored theoretically 
%by \citet[][and references therein]{Bogovalov2015}.
%To our knowledge, closed form formulas for the dispersion relation have not yet been achieved.
%And moreover, these studies focussed on particularly constrained flow regimes 
%of extremely high rotational frequency.

In Section~\ref{sec:modeq} we introduce the compressible Euler equations in a rotating frame of reference as our governing equations. In order to achieve analytical progress, we will focus in Section~\ref{sec:shallow} on a two-dimensional shallow fluid layer which allows for a simplified treatment in terms of perturbation theory. Three different asymptotic regimes distinguished by the angular frequency of the centrifuge
will be studied and evaluated with regard to their similarity to the atmosphere.
In Section~\ref{sec:wkb} the shallow-fluid assumption will be abandoned and three-dimensional waves that extend deep into the interior of the centrifuge are analyzed by means of Wentzel-Kramers-Brillouin theory.
A conclusion, that contains a summary and discussions on the excitation and measurement of waves in centrifuges, will be given in Section~\ref{sec:conclusion}.

\section{The model equations for waves in a gas centrifuge}
\label{sec:modeq}

\begin{figure}
	\centering
	\includegraphics[scale=0.3]{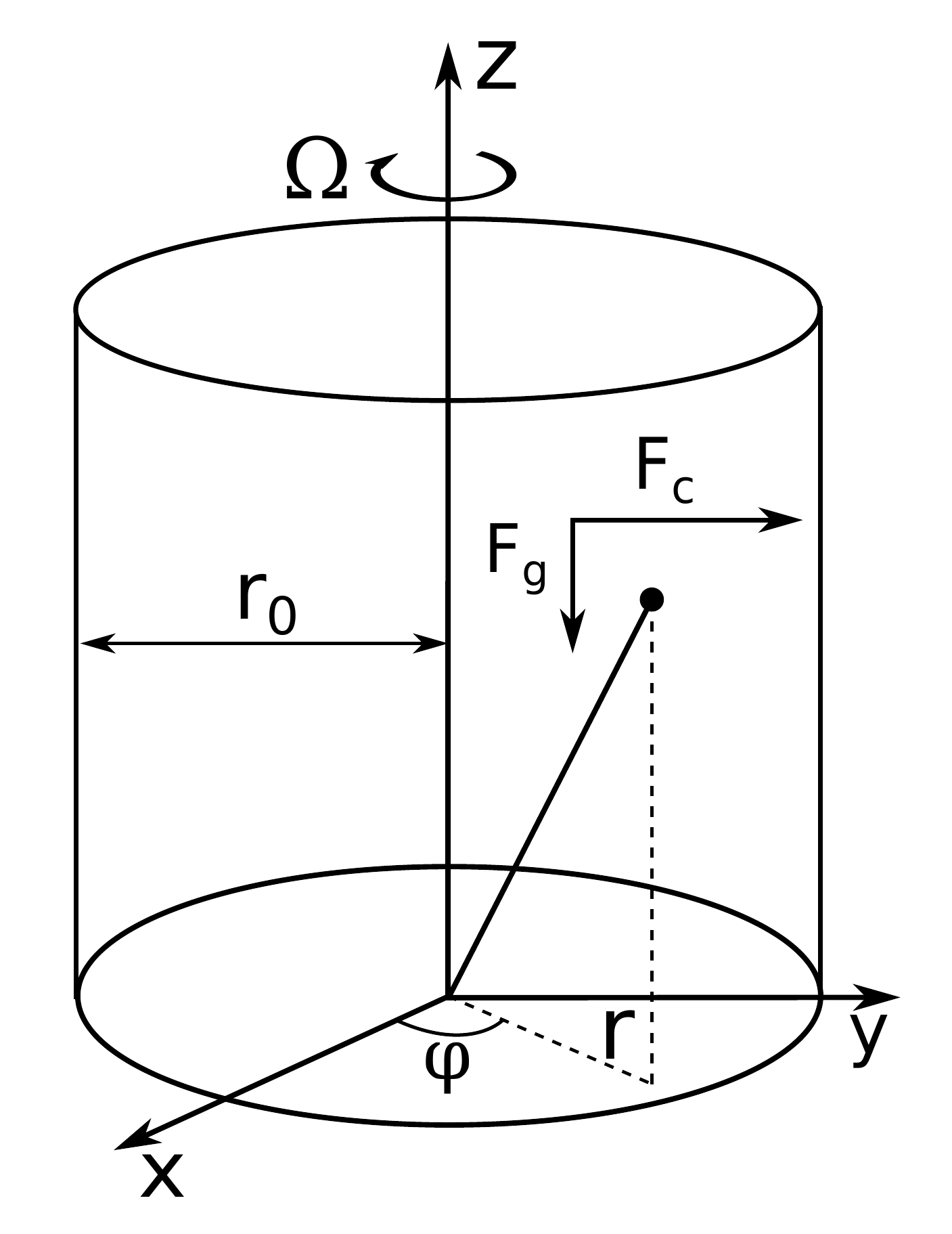}
	\caption{The geometry of the gas centrifuge. 
		Where $\boldsymbol{F_g}$ and $\boldsymbol{F_c}$ represent
	the force of gravity and centrifugal force, respectively.}
	\label{fig:sketch_centrifuge}
\end{figure}

Let us consider an ideal gas in a rotating centrifuge of radius $r_0$ and angular frequency $\Omega$ 
(see Figure~\ref{fig:sketch_centrifuge}). 
%
%\begin{figure}[h]
%	\centering
%	\begin{minipage}[b]{0.3\textwidth}
%		a)\\
%		\includegraphics[width=\textwidth]{sketch_annulus1.png}
%	\end{minipage}
%	\hspace{4mm}
%	\begin{minipage}[b]{0.3\textwidth}
%		b)\\
%		\includegraphics[width=\textwidth]{sketch_cylinder.png}
%	\end{minipage}
%	\caption{The geometry of the gas centrifuge.}
%	\label{fig:sketch_centrifuge}
%\end{figure}
%
The most natural choice for our model are cylindrical coordinates $(r,\varphi,z)$ which denote the radial coordinate, azimuthal angle and axial coordinate, respectively. The dynamical state of the fluid is completely determined by the radial velocity $u$, the azimuthal velocity $v$, the axial velocity $w$, potential temperature $\theta$ and Exner pressure $\pi$. 
The latter two are widely used thermodynamic variables in the atmospheric sciences and 
can be derived from the canonical thermodynamic variables pressure $p$ and temperature $T$ by
\begin{align}
	\label{eq:def_theta_pi}
	\pi=(p/p_{0})^{R/c_p},\quad\theta=T/\pi
\end{align}
where $p_0$ represents a reference pressure at $r_0$. The thermodynamic properties of the ideal gas are specified by the specific gas constant $R$ as well as its specific heat capacities at constant pressure and volume denoted by $c_p$ and $c_v$ that satisfy $R=c_p-c_v$.
Additionally, let us also introduce the dimensionless heat capacities $\hat{c}_p=c_p/R$ and $\hat{c}_v=c_v/R$ such that $1=\hat{c}_p-\hat{c}_v$.
From the theorem of equipartition of energy we know for typical monoatomic gases that $\hat{c}_p=5/2$ and $\hat{c}_v=3/2$, 
whereas for typical diatomic gases we have $\hat{c}_p=7/2$ and $\hat{c}_v=5/2$. 
These values might differ for some molecules and for extremely low or extremely high temperatures.

For later reference we also introduce the equation of state for ideal gases. 
It provides the density $d$ as a function of the thermodynamic state variables,
\begin{align}
	\label{eq:gas_law}
	d= \frac{p}{R T}.
\end{align}

The dynamics of the flow is governed by the compressible, inviscid Euler equations \citep{Achatz2010} in the rotating frame of reference 
expressed in the cylindrical coordinates,
\begin{subequations}
\label{eq:govern}
\begin{align}
	\label{eq:govern_u}
	\frac{\partial u}{\partial t} 
	+u\frac{\partial u}{\partial r} 
	+\frac{v}{r}\frac{\partial u}{\partial \varphi} 
	-\frac{v^2}{r} 
	+w\frac{\partial u}{\partial z}
	+c_p\theta\frac{\partial \pi}{\partial r} &= 
	2\Omega v + r\Omega^2 \\
	\label{eq:govern_v}
	\frac{\partial v}{\partial t} 
	+u \frac{\partial v}{\partial r} 
	+\frac{v}{r}\frac{\partial v}{\partial \varphi} 
	+\frac{v u}{r} 
	+w\frac{\partial v}{\partial z}
	+c_p\frac{\theta }{r}\frac{\partial \pi}{\partial \varphi} &= 
	-2 \Omega u \\
	\label{eq:govern_w}
	\frac{\partial w}{\partial t} 
	+u\frac{\partial w}{\partial r} 
	+\frac{v}{r}\frac{\partial w}{\partial\varphi} 
	+w\frac{\partial w}{\partial z}
	+c_p\theta\frac{\partial \pi}{\partial z}&= -g \\
	\label{eq:govern_theta}
	\frac{\partial \theta }{\partial t} 
	+u\frac{\partial \theta}{\partial r} 
	+\frac{v}{r}\frac{\partial\theta}{\partial\varphi} 
	+w\frac{\partial \theta}{\partial z}&= 0 \\
	\label{eq:govern_pi}
	\frac{\partial \pi}{\partial t} 
	+u \frac{\partial \pi}{\partial r} 
	+\frac{v}{r} \frac{\partial \pi}{\partial \varphi} 
	+w\frac{\partial \pi}{\partial z}
	+\frac{R}{c_v}\pi \left(\frac{1}{r}\frac{\partial(ru)}{\partial r}
	+\frac{1}{r}\frac{\partial v}{\partial \varphi}
	+\frac{\partial w}{\partial z}\right) &= 0.
\end{align}
\end{subequations}
Furthermore, we have assumed that the axial coordinate is parallel to the vertical such that gravity acts on the axial momentum equation where $g$ denotes Earth's gravitational acceleration. 
%We will assume throughout this paper that the centrifugal force is so strong that the gravitational force can be neglected. In fact, $g/(r\omega^2) \sim 10^{-5}$ at $10\, 000 \, \text{RPM}$.

\section{Waves in a shallow stratified layer}
\label{sec:shallow}

In this section we study a simplified version of the governing equations
that allows for analytical progress and comparison with flow regimes of the Earth's atmosphere.
We focus on a shallow two-dimensional layer and ignore gravity for the time being.
The restrictions will be weakened again in Section~\ref{sec:wkb}
where we will use the results from this section to build a comprehensive
three-dimensional wave theory for a particular flow regime.

\subsection{Simplified model and characteristic polynomial}
First, we restrict our analysis to axially homogeneous solutions neglecting gravity
as the centrifugal force is presumably much stronger. 
Second, we apply the shallow-fluid approximation \citep[cf.][]{Vallis2006}. 
Let $\Delta r$ be the thickness of a thin fluid layer at the rim of the centrifuge 
such that $\Delta r\ll r_0$.
We introduce a new coordinate $r'=r-r_0$ and assume that $r' \sim \Delta r$. 
Due to the assumption, we focus on the dynamics close to $r_0$, 
the rim of the centrifuge
ignoring any boundary layer effects.
The governing equations \eqref{eq:govern} take the following form,
%
%, the coordinate $r$ is hence replaced by $r_0$ except where it appears as the argument in the derivative; the metric coefficients become constants,
\begin{subequations}
\label{eq:govern2}
\begin{align}
	\label{eq:govern_u2}
	\frac{\partial u}{\partial t} + u\frac{\partial u}{\partial r'} + \frac{v}{r_0}\frac{\partial u}{\partial \varphi} 
	-\frac{v^2}{r_0} + c_p\theta\frac{\partial \pi}{\partial r'} &= 2\Omega v + r_0\Omega^2 \\
	\label{eq:govern_v2}
	\frac{\partial v}{\partial t} + u \frac{\partial v}{\partial r'} +\frac{v}{r_0}\frac{\partial v}{\partial \varphi} 
	+\frac{v u}{r_0} + c_p \frac{\theta}{r_0}\frac{\partial \pi}{\partial \varphi} &= -2 \Omega u \\
	\label{eq:govern_theta2}
	\frac{\partial \theta }{\partial t} + u\frac{\partial \theta}{\partial r'} +\frac{v}{r_0}\frac{\partial\theta}{\partial\varphi} &= 0 \\
	\label{eq:govern_pi2}
	\frac{\partial \pi}{\partial t} + u \frac{\partial \pi}{\partial r'} + \frac{v}{r_0} \frac{\partial \pi}{\partial \varphi} 
	+ \frac{R}{c_v}\pi \left(\frac{\partial u}{\partial r'}
	+\frac{1}{r_0}\frac{\partial v}{\partial \varphi}\right) &= 0.
\end{align}
\end{subequations}
The shallow-fluid approximation transforms the problem to Eucledian geometry 
as the metric coefficients become constants.

Let us define a background state by establishing a fluid at rest in the rotating frame of reference
which is a solution to the governing equations in the shallow-fluid approximation \eqref{eq:govern2}.
This state is determined by the rigid body rotation of the stratified gas,
\begin{align}
\begin{pmatrix}
		u\\ v\\ \theta\\ \pi
\end{pmatrix}(r',\varphi,t)
=\begin{pmatrix}
	0\\ 0\\ \Theta\\ \Pi
\end{pmatrix}(r').
\end{align}
The background fulfills the radial momentum equation \eqref{eq:govern_u2} 
which is equivalent to the hydrostatic balance in the atmosphere,
\begin{align}
	\label{eq:hydrostatic_balance_rotating}
	c_p \Theta\frac{d\Pi}{dr'} = r_0 \Omega^2.
\end{align}
If we additionally assume that the background state is isothermal, 
i.e. the unperturbed gas is in thermodynamic equilibrium,
\begin{align}
	\Pi\Theta=T_0=\mathrm{const.}
\end{align}
we can easily solve \eqref{eq:hydrostatic_balance_rotating} for the background Exner pressure and potential temperature applying the boundary conditions $\Pi(0)=\Pi_0$ and $\Theta(0)=\Theta_0$,
\begin{subequations}
\begin{align}
%	\Pi(r)=\Pi_0\,e^{r/r_\theta},\quad
	\Pi(r')=\Pi_0\,\exp\left(\frac{r'}{r_\theta}\right),\quad
%	\Theta(r)=\Theta_0\,e^{-r/r_\theta}.
	\Theta(r')=\Theta_0\,\exp\left(-\frac{r'}{r_\theta}\right).
\end{align}
\end{subequations}
Utilizing the equation of state for ideal gases \eqref{eq:gas_law}, we also obtain the background density,
\begin{align}
%	D(r)=D_0\,e^{r/r_d}.
	D(r')=D_0\,\exp\left(\frac{r'}{r_d}\right).
\end{align}
Here, we defined in analogy to the scale heights in the atmosphere the potential temperature and the density scale radius by
\begin{align}
	r_\theta=\frac{c_pT_0}{r_0\Omega^2},\quad r_d=r_\theta/\hat{c}_p.
\end{align}
Notice that the stratified gas in the centrifuge exhibits a similar exponential behavior as the hydrostatic atmosphere.
We will see in the next section that this is only true in the shallow-fluid approximation. 
However, even when taking the curved geometry into account, 
the stratification in the centrifuge can still be computed analytically.

In order to study waves as perturbations to our stratified gas we insert the ansatz
\begin{subequations}
\begin{align}
	u(r',\varphi,t)&=u'(r',\varphi,t)\\
	v(r',\varphi,t)&=v'(r',\varphi,t)\\
	\theta(r',\varphi,t)&=\Theta(r')+\theta'(r',\varphi,t)\\
	\pi(r',\varphi,t)&=\Pi(r')+\pi'(r',\varphi,t)
\end{align}
\end{subequations}
into the simplified governing equations \eqref{eq:govern2} and linearize around the background state.
It will turn out to be useful to rescale the perturbation field, 
\begin{subequations}
\label{eq:rescale_prognostic}
\begin{align}
	\tilde{u}(r',\varphi,t)&=D(r')^{1/2}\,u'(r',\varphi,t)\\
	\tilde{v}(r',\varphi,t)&=D(r')^{1/2}\,v'(r',\varphi,t)\\
	\tilde{\theta}(r',\varphi,t)&=\hat{c}_v^{1/2}C_s\,D(r')^{1/2}\,\frac{\theta'(r',\varphi,t)}{\Theta(r')}\\
	\tilde{\pi}(r',\varphi,t)&=\hat{c}_vC_s\,D(r')^{1/2}\,\frac{\pi'(r',\varphi,t)}{\Pi(r')}
\end{align}
\end{subequations}
where $C_s=\sqrt{\gamma RT_0}$ denotes the usual speed of sound 
and $\gamma=c_p/c_v=\hat{c}_p/\hat{c}_v$ is the heat capacity ratio.
The variable transformation \eqref{eq:rescale_prognostic} standardizes 
the units of the prognostic variables---they all have gotten the dimension square root of energy. 
Moreover, it anticipates that the amplitudes grow exponentially away from the rim 
into the interior, analogously to the altitudinal amplification in the atmosphere \citep{Durran1989}.
The resulting linear system for the transformed prognostic variables of the perturbation reads
\begin{subequations}
\label{eq:lin_govern}
\begin{align}
	\frac{\partial \tilde{u}}{\partial t}+C_s\,\frac{\partial\tilde{\pi}}{\partial r'}-C_s\eta\,\tilde{\pi}
	+N_0\,\tilde{\theta}-2\Omega\,\tilde{v}&=0\\
	\frac{\partial\tilde{v}}{\partial t}+\frac{C_s}{r_0}\,\frac{\partial\tilde{\pi}}{\partial\varphi}
	+2\Omega\,\tilde{u}&=0\\
	\frac{\partial\tilde{\theta}}{\partial t}-N_0\,\tilde{u}&=0\\
	\frac{\partial\tilde{\pi}}{\partial t}+C_s\,\frac{\partial \tilde{u}}{\partial r'}
	+C_s\eta\,\tilde{u}+\frac{C_s}{r_0}\,\frac{\partial\tilde{v}}{\partial\varphi}&=0.
\end{align}
\end{subequations}
where we defined the reference Brunt-Väisälä frequency as 
\begin{align}
	N_0=\sqrt{-\frac{r_0\Omega^2}{\Theta}\frac{d\Theta}{dr'}}=\frac{r_0\Omega^2}{\sqrt{c_pT_0}}.
\end{align}
An additional parameter appears that is defined by
\begin{align}
	\eta=\frac{1}{2r_d}-\frac{1}{r_\theta}.
\end{align}
%It is the inverse of the scale radius of the combined background quantity $D^{1/2}(r')\Theta(r')$.
A similar term occurs in \cite{Durran1989}.
It can readily be show that \eqref{eq:lin_govern} conserves the total energy density $\tilde{u}^2+\tilde{v}^2+\tilde{\theta}^2+\tilde{\pi}^2$ which consists of the sum of kinetic, potential and internal energy density \citep[cf.][Eq. (2.16)]{Achatz2010}. 
In the light of this consideration we can interpret $C_s\eta$ as an energy exchange rate for the transformation between kinetic and internal energy which is associated with the compressibility of the gas.

Note that the emerging linear system of partial differential equations possesses only constant coefficients.
Therefore, we apply a plane wave ansatz for the perturbation field
\begin{align}
	\begin{pmatrix}
		\tilde{u}\\
		\tilde{v}\\ 
		\tilde{\theta}\\
		\tilde{\pi}\\
	\end{pmatrix}(r',\varphi,t)=  
	\begin{pmatrix}
		a_u\\
		a_v\\
		a_\theta\\
		a_\pi 
	\end{pmatrix}
	\exp( \mathrm{i}m r' + \mathrm{i}\kappa\varphi-\mathrm{i}\omega t).
	\label{eq:plane_wave_ansatz}
\end{align}
The azimuthal wavenumber is denoted by $\kappa$. To ensure periodic waves along the azimuth it is required to be an integer. 
We call $m$ the radial wave number which is in the reals as we ignore boundary conditions for the time being.
And the wave frequency is $\omega$.
%radially modulated plane wave ansatz
%\begin{align}
%	\begin{pmatrix}
%		u'\\
%		v'\\ 
%		\theta'\\
%		\pi'
%	\end{pmatrix}(r,\varphi,t)=  
%	\begin{pmatrix}
%		\hat{u}\\
%		\hat{v}\\
%		\hat{\theta}\\
%		\hat{\pi} 
%	\end{pmatrix}(r)\,
%	\exp(-i\omega t+imr+ik\varphi)
%	\label{eq:plane_wave_ansatz}
%\end{align}
Substituting the plane wave ansatz in \eqref{eq:lin_govern} we obtain an algebraic equation for the amplitudes
\begin{align}
    \begin{pmatrix}
	    -\mathrm{i}\omega & -2\Omega & N_0 & C_s (\mathrm{i}m-\eta)\\
	    2\Omega & -\mathrm{i}\omega & 0 & \mathrm{i}C_s\kappa/r_0 \\
	    -N_0 & 0 & -\mathrm{i}\omega & 0 \\
	    C_s(\mathrm{i}m + \eta) & \mathrm{i}C_s \kappa/r_0 & 0 & -\mathrm{i}\omega
    \end{pmatrix}
    \begin{pmatrix}
	    a_u \\
	    a_v \\
	    a_\theta \\
	    a_\pi
    \end{pmatrix}
    =0.
\label{eq:matrix_equation}
\end{align}

%\begin{align}
%    \begin{pmatrix}
%	    -i\omega & -2\Omega & N & C_s (im-\eta)\\
%	    2\Omega & -i\omega & 0 & iC_sk/r_0 \\
%	    -N & 0 & -i\omega & 0 \\
%	    C_s(im + \eta) & iC_s k/r_0 & 0 & -i\omega
%    \end{pmatrix}
%    \begin{pmatrix}
%	    \hat{u} e^{r/(2\kappa r_\theta)}\\
%	    \hat{v} e^{r/(2\kappa r_\theta)}\\
%	    \hat{\theta} \frac{r_0\Omega^2}{N\Theta}\,e^{r/(2\kappa r_\theta)}\\
%	    \hat{\pi}\frac{c_p\Theta}{C_s}\,e^{r/(2\kappa r_\theta)}
%    \end{pmatrix}
%    =0
%\label{matrix_equation}
%\end{align}

This linear system of equations can be written in the form $\mathsfbi{M}\,\boldsymbol{a}= \mathrm{i}\omega\,\boldsymbol{a}$ 
with $\mathsfbi{M}$ being the system matrix and $\boldsymbol{a}$ a vector containing the amplitudes of the perturbation flow variables.
Therefore, we are actually solving an eigenvalue problem.
The eigenvalues $\mathrm{i}\omega$ of $\mathsfbi{M}$ are given as the roots of its characteristic polynomial
\begin{align}
	\omega^4-\left(C_s^2\eta^2+N_0^2+4\Omega^2+C_s^2 \frac{\kappa^2}{r_0^2}+C_s^2m^2\right)\omega^2
	-4C_s^2\eta\Omega\frac{\kappa}{r_0}\omega+C_s^2N_0^2\frac{\kappa^2}{r_0^2}=0 
\end{align}
which is of fourth order.
Note that the characteristic polynomial is almost the same as for perturbations of the Earth's hydrostatic atmosphere
where the coefficients would be defined accordingly and the horizontal wavenumber would be $k=\kappa/r_0$.
In comparison to the atmosphere two additional terms arise for the gas centrifuge,
the linear term in $\omega$ and the term $4\Omega^2$ in the coefficient of the quadratic term.
Both extra terms are due to the Coriolis force that acts in the same plane as the centrifugal force
in contrast to the atmosphere where Coriolis and gravitational force are orthogonal.
Note that the latter statement is true only in the traditional approximation where the Coriolis force is projected onto the horizontal plane \citep[cf.][]{Vallis2006}.

Despite the fact that quartic polynomials possess analytic roots,
we will not gain much by computing them explicitly as they are tedious.
Nevertheless, useful insight about the properties of the eigenvalues is provided directly by the structure of the matrix. 
Notice that $\mathsfbi{M}$ is skew-Hermitian, 
i.e. $\mathsfbi{M}^\mathrm{H}=-\mathsfbi{M}$
where $\mathrm{H}$ denotes the conjugate transpose.
This property implies that the matrix has only imaginary eigenvalues, 
further implying that $\omega$ must be real. 
Thus, we conclude that the stratified gas at rest in the centrifuge is stable
since all solutions to the governing equations as given by \eqref{eq:plane_wave_ansatz} remain bounded as $t\rightarrow\infty$.

\subsection{Non-dimensionalization of the characteristic polynomial}

With the aid of asymptotic analysis we can find approximations to the roots of the characteristic polynomial 
and identify regimes that are similar to the atmosphere.
For this task we need to non-dimensionalize the equations.
A convenient choice for dimensionless frequency and radial wavenumber is provided by
\begin{align}
	\label{eq:def_sigma}
	\sigma=\omega/N_0,\quad\mu=r_0m.
\end{align}
Then, the dimensionless characteristic polynomial reads
\begin{align}
	\hat{c}_v\,\sigma^4-\left[\hat{c}_p^2/4+4\hat{c}_vq+q^2\left(\kappa^2+\mu^2\right)\right]\,\sigma^2
	-4\hat{\eta}\,q^{3/2}\kappa\,\sigma+q^2\kappa^2=0 
\label{eq:nondim_charpol}
\end{align}
where we introduced the parameters
\begin{align}
	\label{eq:def_q}
	q=\frac{r_\theta}{r_0}=\frac{\Omega^2}{N_0^2}=\frac{c_pT_0}{r_0^2\Omega^2},\quad \hat{\eta}=\hat{c}_p/2-1.
\end{align}
The parameter $\hat{\eta}$ is determined by the properties of the working gas and is considered to be a constant.
The variable $q$ on the other hand might vary by several orders of magnitude depending on the choice of angular frequency $\Omega$, say.
Hereinafter we will refer to it as non-dimensional scale radius.
In order to employ asymptotic analysis a universally small number is needed that helps to separate scaling regimes of interest.
An immediate scale separation parameter presents itself in terms 
of the ratio of fluid layer thickness $\Delta r$ and the radius of the centrifuge
which is naturally a small number due to the shallow-fluid approximation, so we define
\begin{align}
	\varepsilon=\frac{\Delta r}{r_0}\ll 1.
\end{align}
In order to obtain meaningful wave solutions, 
at least one radial wavelength $\lambda_0$ must fit into
the thin layer, 
i.e. $\lambda_0\sim \Delta r$, and hence $\lambda_0/r_0=\mathit{O}(\varepsilon)$.
Based on this restriction, we can write
\begin{align}
	\label{eq:def_mu}
	\mu=\varepsilon^{-1}M
\end{align}
and require that $M=\mathit{O}(1)$ as $\varepsilon\rightarrow 0$
which is our first step towards a distinguished limit.

%An immediate scale separation parameter presents itself in terms 
%of the ratio of the reference radial wavelength and the radius of the centrifuge, 
%\begin{align}
%	0<\varepsilon=\lambda_0/r_0\ll 1.
%\end{align} 
%This ratio has to be small to meet the shallow-fluid assumption;
%We need that at least one radial wavelength fits in the thin fluid layer.
%Based on this observation, we can write
%\begin{align}
%	\label{eq:def_mu}
%	\mu=\varepsilon^{-1}M
%\end{align}
%and require that $M=\mathit{O}(1)$ as $\varepsilon\rightarrow 0$
%which is our first step towards a distinguished limit.
In the following we present systematically three asymptotic regimes 
in terms of their distinguished limits 
that are of interest for the atmospheric sciences.
For an overview of the different regimes a diagram, 
that also clarifies our naming conventions and abbreviations,
will be presented in Section~\ref{sec:compare}.

%\section{Distiguished limits and asymptotic regimes}
\subsection{Asymptotic regime of low angular frequency}
\label{sec:low}

In this section we will study the particular asymptotic regime 
where the non-dimensional scale radius fulfills $1\ll q=\mathit{O}(\varepsilon^{-1})$, i.e. it is very large. 
If we keep the centrifuge's radius and the temperature constant, then $q$ increases according to \eqref{eq:def_q} when we decrease the angular frequency.
Since the latter is probably the easiest parameter to adjust in an experimental gas centrifuge device, 
we will coin the several asymptotic regimes in terms of their angular frequencies.
Let us assume for the moment that we used air as the working gas and 
that $T_0=300$\,K and $r_0=0.5$\,m to gain some intuition 
about the regime of low angular frequency.
For $\varepsilon=0.01$ the rotational frequency would be approximately $1000$\,RPM.

%\subsubsection{Dispersion relation for isotropic wave field}

Let us consider the following distinguished limit,
\begin{align}
	\label{eq:dist_lim_low_iso}
	q=\varepsilon^{-1}Q,\quad \kappa=\lceil\varepsilon^{-1}K\rceil,\quad Q,K=\mathit{O}(1)\text{ as } \varepsilon\rightarrow 0
\end{align}
where $\lceil\cdot\rceil$ denotes the ceiling function ensuring that $\kappa$ is an integer and hence the solution azimuthally periodic.
Since the radial wavenumber $\mu=\mathit{O}(\varepsilon^{-1})$ has the same order as the azimuthal wavenumber
such that $\kappa/\mu=\mathit{O}(1)$,
we identify this regime to be isotropic, i.e. no direction is to be favored.
The interested reader finds a detailed analysis of an anisotropic regime, where the azimuthal wavenumber is much shorter than the radial, in the Appendix~\ref{app:low}.
Notice that the scale of the horizontal wavelength and hence the wavenumber 
may be practically determined by the boundary condition at the rim.
When we insert the distinguished limit into \eqref{eq:nondim_charpol}, 
we obtain a polynomial in $\sigma$ whose coefficients depend on constant $\mathit{O}(1)$-parameters and $\varepsilon$,
\begin{align}
	\varepsilon^4\hat{c}_v\,\sigma^4-\left[\varepsilon^4\hat{c}_p^2/4
	+4\varepsilon^3\hat{c}_vQ+Q^2\left(K^2+M^2\right)\right]\,\sigma^2
	-4\varepsilon^{3/2}\hat{\eta}Q^{3/2}K\,&\sigma+Q^2K^2=0 .
\label{eq:bgw_charpol1}
\end{align}
We pass to the limit $\varepsilon\rightarrow 0$ giving us 
\begin{align}
	-Q^2\left(K^2+M^2\right)\,&{\sigma^{(0)}}^2+Q^2K^2=0.
\label{eq:leading_order_bwg}
\end{align}
The solution provides two distinguished roots
\begin{align}
	{\sigma^{(0)}}^2=\frac{K^2}{K^2+M^2}.
\end{align}
When we substitute the dimensionless by the dimensional variables according to the scaling assumptions 
\eqref{eq:dist_lim_low_iso} and the definitions \eqref{eq:def_sigma}, 
\eqref{eq:def_q} and \eqref{eq:def_mu} we arrive at a redimensionalization of the roots,
\begin{align}
	\frac{\omega^2}{N_0^2}\approx\frac{\kappa^2}{\kappa^2+r_0^2m^2}
\end{align}
or equivalently
\begin{align}
	\label{eq:dispersion_bgw}
	\omega_\mathrm{BGW}^2\approx\frac{N_0^2k^2}{k^2+m^2},\quad k=\kappa/r_0.
\end{align}
This is the exact same dispersion relation as for atmospheric non-hydrostatic gravity waves.
As we assume in this regime that the scale radius tends to infinity as stated by \eqref{eq:def_q},
the background density becomes constant which is equivalent to the Boussinesq approximation.
Hence, the stratification is weak and effects due to the compressibility of the gas become negligible.
The resulting waves are the same as the wave solutions to the Boussinesq equations (BGW).
Consequently, the waves do not encounter radial amplification---the centrifuge equivalent of altitudinal amplification in the atmosphere---but approximately constant amplitudes in the radial direction.
This regime is hence more similar to internal waves in water, 
which is stratified due to salinity for example,
than to atmospheric gravity waves.

Notice that the leading order equation \eqref{eq:leading_order_bwg} provided only two roots. Two others tend to infinity.
Therefore, we face a singular perturbation problem. 
It can be solved by the method of dominant balance \citep{Bender1999} that provides an appropriate rescaling,
\begin{align}
	\sigma=\varepsilon^{-2}\Sigma.
\end{align}
Next, we substitute the rescaled frequency in the dimensionless characteristic polynomial,
\begin{align}
	\hat{c}_v\,\Sigma^4-\left[\varepsilon^4\hat{c}_p^2/4
	+4\varepsilon^3\hat{c}_vQ+Q^2\left(K^2+M^2\right)\right]\,\Sigma^2
	-4\varepsilon^{7/2}\hat{\eta}Q^{3/2}K\,&\Sigma+\varepsilon^4Q^2K^2=0. 
\label{eq:bgw_charpol2}
\end{align}
As $\varepsilon\rightarrow 0$ we obtain the leading order equation
\begin{align}
	\hat{c}_v\,{\Sigma^{(0)}}^4-Q^2\left(K^2+M^2\right)\,{\Sigma^{(0)}}^2=0
\label{}
\end{align}
which is solved by four roots.
Two of them are given by
\begin{align}
	{\Sigma^{(0)}}^2=\hat{c}_v^{-1}Q^2\left(K^2+M^2\right).
\end{align}
The other two roots are zero. They correspond to the two roots 
that we have already identified in the original scaling.
Let us redimensionalize the two non-vanishing roots equivalently to \eqref{eq:dispersion_bgw},
\begin{align}
	\omega_\mathrm{A}^2\approx C_s^2\left(k^2+m^2\right),\quad k=\kappa/r_0.
\end{align}
We notice that this is the exact same dispersion relation as for usual acoustic waves (A).

%\subsubsection{lfngljdsfhg}

Remarkably, the two modes, that we found in both regimes, i.e. the internal waves similar to Boussinesq gravity waves and the acoustic waves, exhibit an exceedingly strong scale separation of $\mathit{O}(\varepsilon^2)$.
As a consequence of this scale separation the phase and group velocities of these two modes 
being defined, respectively, by
\begin{align}
	\label{eq:phase_group_velo}
	\boldsymbol{c}_p=\frac{\boldsymbol{k}\omega}{\|\boldsymbol{k}\|^2},~
	\boldsymbol{c}_g=\frac{\partial\omega}{\partial\boldsymbol{k}}
\end{align}
where
\begin{align}
	\boldsymbol{k}=\begin{pmatrix}
		k\\ m
	\end{pmatrix},~
	\|\boldsymbol{k}\|=\sqrt{k^2+m^2}
\end{align}
will also differ by the same order in the scale separation parameter,
i.e. the acoustic will propagate much faster than the internal wave modes.

\subsection{Asymptotic regime of intermediate angular frequency}
\label{sec:intermediate}

For this regime, we assume that the non-dimensional scale radius fulfills $q=\mathit{O}(1)$ 
which implies that the potential temperature scale radius 
is of the same order of magnitude as the radius of the centrifuge.
To gain an intuition for this regime with air as the working gas, 
let us assume for the moment that $T_0=300$\,K and $r_0=0.5$\,m.
Then, the rotational frequency would be around 10\,000\,RPM. 

%\subsubsection*{Isotropic wave field}

If we additionally assume isotropy in the wave field, we arrive at the following scaling assumptions
\begin{align}
	\label{eq:scaling_intermediate}
	q= Q,\quad \kappa=\lceil\varepsilon^{-1}K\rceil,\quad Q,K=\mathit{O}(1)\text{ as } \varepsilon\rightarrow 0.
\end{align}
The results for anisotropic scaling can be found in the Appendix~\ref{app:inter}.
In the limit $\varepsilon\rightarrow 0$ we find two roots at the leading order
%\begin{align}
%	\hat{c}_v^{-1}Q^2\left(K^2+M^2\right)\,{\sigma^{(0)}}^2
%	+\hat{c}_v^{-1}Q^2K^2 = 0 
%\end{align}
\begin{align}
	{\sigma^{(0)}}^2=\frac{K^2}{K^2+M^2}
\end{align}
which become
\begin{align}
	\label{eq:gw_dispersion}
	\omega^2_\mathrm{GW}\approx\frac{N_0^2k^2}{k^2+m^2}
\end{align}
when we redimensionalize the variables.
This formula is identical to the dispersion relation of non-hydrostatic atmospheric gravity waves (GW).
Note that in contrast to the regime of low angular frequency the scale radius remains finite 
in the asymptotic limit.
The background density varies by a factor of approximately $e$---the Euler constant---in the domain of interest.
The waves, indeed, encounter radial amplification
and thereby closely resemble atmospheric gravity waves.
In comparison to the regime of low angular frequency resembling the dynamics of the Boussinesq equations, the internal waves in the intermediate regime are, hence, equivalent to the wave solutions of the pseudo-incompressible equations \cite{Durran1989}. The latter depict an improvement of the anelastic equations of \cite{Lipps1982}.
\cite{Achatz2010} concluded that the anelastic equations are only valid for gravity waves when the potential temperature scale height is much larger than the Exner pressure scale height. In our scenario the corresponding scale radii are of the same order and hence the anelastic equations would not be applicable.

Analogous to the regime of low angular frequency, we find only two roots for the leading order polynomial.
The other two roots are recovered by the rescaling
\begin{align}
	\sigma=\varepsilon^{-1}\Sigma.
\end{align}
To leading order of the rescaled characteristic polynomial
two non-vanishing roots are obtained
\begin{align}
	{\Sigma^{(0)}}^2=\hat{c}_v^{-1}Q^2\left(K^2+M^2\right).
\end{align}
In the next step, we replace the dimensionless variables by their dimensional counterparts to derive
\begin{align}
	\omega^2_\mathrm{A}\approx C_s^2\left(k^2+m^2\right)
\end{align}
which resembles the dispersion relation of acoustic waves (A).

In conclusion, we obtain a clear scale separation between acoustic waves and internal waves of $\mathit{O}(\varepsilon)$ 
which exhibit to leading order 
the exact same dispersion relations as their atmospheric siblings.
This regime is of particular interest with regard to using the gas centrifuge as an experimental device to study atmospheric waves
because the waves additionally experience radial amplification, 
in contrast to the regime of low angular frequency since the scale radius is finite.
Due to its relevance to the atmosphere, we want to study the regime of intermediate angular frequency in more detail.
In our leading order results, any effect due to the Coriolis force vanishes.
First, we want to find out at which order and how the Coriolis force alters the wave properties. 
And second, we want to relax the shallow-fluid approximation. 
We dedicate the Section~\ref{sec:wkb} to the latter endeavor.

\subsubsection{Higher-order correction of the dispersion relation due to the Coriolis force}

In order to gain insight into the effect of the Coriolis force we expand the frequency by means of $\varepsilon$,
\begin{align}
	\sigma={\sigma^{(0)}}+\varepsilon{\sigma^{(1)}}+\mathit{O}(\varepsilon^2).
\end{align}

Inserting the ansatz into \eqref{eq:nondim_charpol} and collecting terms in powers of $\varepsilon$, we find at the next order
\begin{align}
	{\sigma^{(1)}}=-2\hat{\eta}Q^{-1/2}\frac{K}{K^2+M^2}.
\end{align}
The influence of the Coriolis force appears as a higher order correction to the leading order solution.
As long as $\varepsilon$ is sufficiently small, i.e. the radial wavelength sufficiently short in comparison to the scale radius,
the leverage of the Coriolis force is negligible.
When we combine the next-order with the leading-order result and redimensionalize as before, we get
\begin{align}
	\omega_\mathrm{GW}\approx\pm\underbrace{\frac{N_0k}{\sqrt{k^2+m^2}}}_{\mathit{O(N_0)}}
	-\underbrace{\frac{2\eta\Omega k}{k^2+m^2}}_{\mathit{O(\varepsilon N_0)}}.
\end{align}

\begin{figure}
	\centering
	\includegraphics[scale=0.8]{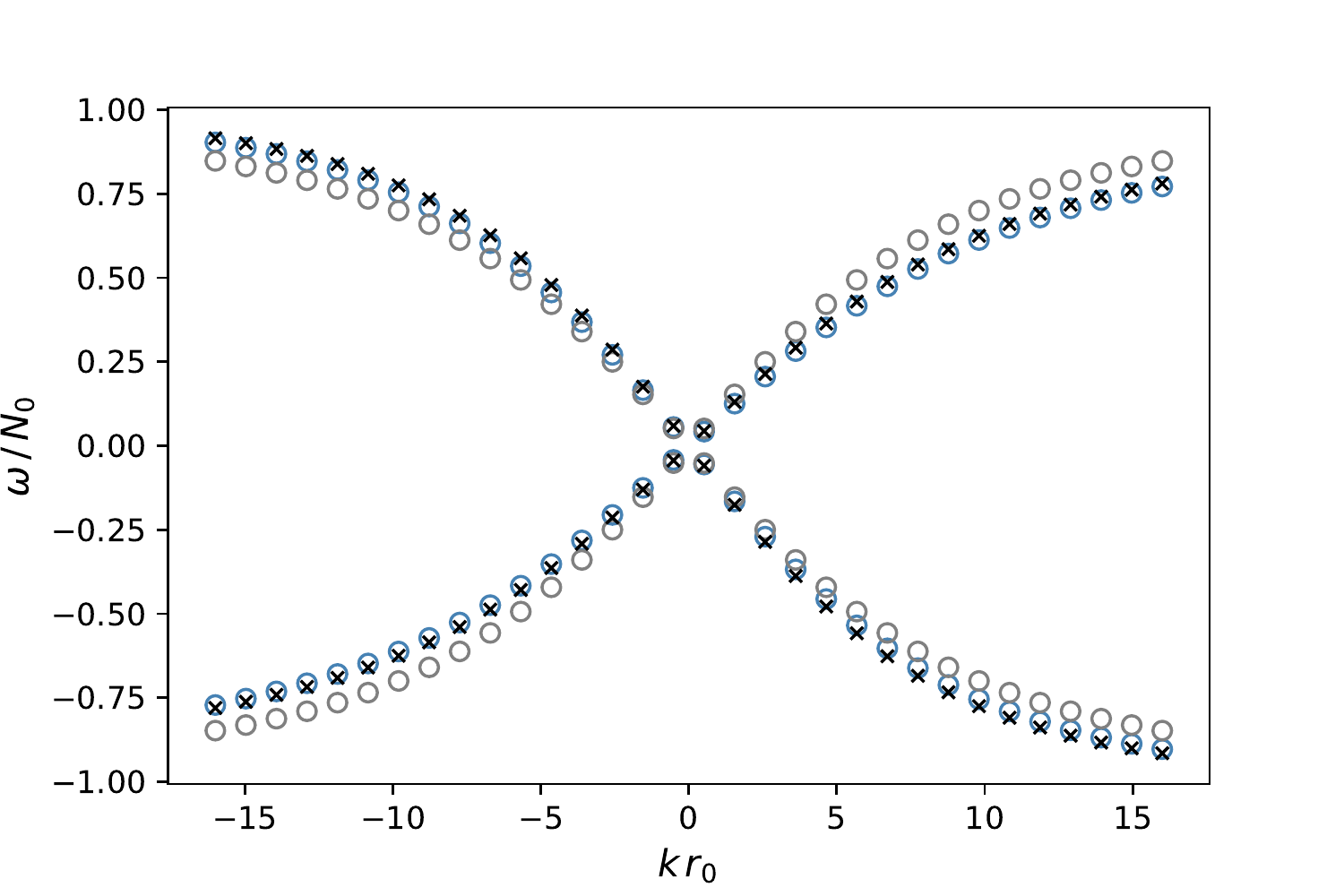}
	\caption{
		Dispersion relation in the regime of intermediate angular frequency. 
		Leading-order asymptotic solution (gray circles),
		asymptotic solution with higher-order correction (blue circles)  
		and exact solution (black crosses) for $m\,r_0=10$ and $\varepsilon=0.1$.
	}
	\label{fig:dispersion}
\end{figure}

Let us briefly elaborate on the properties of the higher-order correction due to the Coriolis force. 
For a given $m$ the unperturbed frequency as a function of $k$ is symmetric with respect to the axis $k=0$,
i.e. there is no difference in frequency and therefore in phase speed between waves propagating along the azimuth or against it.
The higher-order correction breaks this mirror symmetry 
as for positive $k$ the correction is always negative and for negative $k$ vice versa.
There is no sign switch selecting a branch of the correction for the dispersion relation.
An illustration of this asymmetry is given in Figure~\ref{fig:dispersion}. 
Consequently, waves with the same absolute azimuthal and radial wavenumbers
have different frequencies as well as phase speeds and therefore group velocities (cf. Equations~\ref{eq:phase_group_velo}) depending 
on whether they travel with or against the rotation of the centrifuge.
However, this effect is minuscule in the regime at hand.
We plotted the leading-order dispersion relation along with its higher-order correction and the exact solution.
Notice that the corrected asymptotic and the exact solution align almost perfectly.

We also studied the asymptotic correction due to the Coriolis force for the other two regimes.
For the sake of brevity, the details of these calculations are not shown here.
It was found that the effect on the dispersion by the Coriolis force 
is always at least one order higher in $\varepsilon$ than the leading-order solution.

\subsubsection{Polarization and order relations}

In addition to their dispersion relations waves are typically characterized by their polarization properties.
Hence, we also want to study the polarization relations of this particularly interesting regime.
To this end we rewrite the system matrix of \eqref{eq:matrix_equation} in dimensionless form with the distinguished limit of intermediate angular frequency given by \eqref{eq:scaling_intermediate}, so
%\begin{align}
%	\hat{\mathsfbi{M}}=
%	\hat{c}_v^{-1/2}
%	\begin{pmatrix}
%		0& -2\hat{c}_v^{1/2}Q^{-1/2}& \hat{c}_v^{1/2}& i\varepsilon^{-1}QM-\hat{\eta}\\
%		2\hat{c}_v^{1/2}Q^{-1/2}& 0& 0& i\varepsilon^{-1}QK\\
%		-\hat{c}_v^{1/2}& 0& 0& 0\\
%		i\varepsilon^{-1}QM+\hat{\eta}& i\varepsilon^{-1}QK& 0& 0.
%	\end{pmatrix}
%\end{align}
\begin{align}
	\hat{\mathsfbi{M}}=
	\begin{pmatrix}
		0& -2Q^{-1/2}& 1& \hat{c}_v^{-1/2}\left[\mathrm{i}\varepsilon^{-1}QM-\hat{\eta}\right]\\
		2Q^{-1/2}& 0& 0& \mathrm{i}\varepsilon^{-1}\hat{c}_v^{-1/2}QK\\
		-1& 0& 0& 0\\
		\hat{c}_v^{-1/2}\left[\mathrm{i}\varepsilon^{-1}QM+\hat{\eta}\right]& \mathrm{i}\varepsilon^{-1}\hat{c}_v^{-1/2}QK& 0& 0
	\end{pmatrix}.
\end{align}
We notice that the matrix can be decomposed in terms of powers of $\varepsilon$ into 
\begin{align}
	\hat{\mathsfbi{M}}=\varepsilon^{-1}\hat{\mathsfbi{M}}^{(-1)}+\hat{\mathsfbi{M}}^{(0)}
\end{align}
which we exploit to solve the eigenvalue problem $\left(\hat{\mathsfbi{M}}-\mathrm{i}\sigma\right)\boldsymbol{a}=0$.
We already derived the eigenvalues asymptotically by application of perturbation theory on the characteristic polynomial. The eigenvectors represent the polarization properties. In order to find an asymptotic solution for them we expand
\begin{align}
	\boldsymbol{a}=\boldsymbol{a}^{(0)}+\varepsilon\,\boldsymbol{a}^{(1)}+\mathit{O}\left(\varepsilon^2\right).
\end{align}
Inserting the ansatz and collecting terms by powers of $\varepsilon$ we find in the leading order
\begin{align}
	\hat{\mathsfbi{M}}^{(-1)}\boldsymbol{a}^{(0)}=0.
\end{align}
We solve this linear system of equations by expressing the unknowns in terms of the amplitude of the radial velocity giving the polarization relations
\begin{align}
	a_\pi^{(0)}=0,\quad a_v^{(0)}=-\frac{M}{K}a_u^{(0)}.
\end{align}
Hence, to zero-order the amplitude of the rescaled Exner pressure perturbation vanishes.
At the next order we obtain
\begin{align}
	\hat{\mathsfbi{M}}^{(-1)}\boldsymbol{a}^{(1)}+\left(\hat{\mathsfbi{M}}^{(0)}-\mathrm{i}\sigma^{(0)}\right)\boldsymbol{a}^{(0)}=0.
\end{align}
This linear system provides us with the polarization relation for the zero-order rescaled potential temperature 
and eventually a first-order expression for the rescaled Exner pressure
\begin{align}
	a_\theta^{(0)}=\mathrm{i}\frac{1}{\sigma^{(0)}}a_u^{(0)},\quad a_\pi^{(1)}=\frac{2\mathrm{i}Q^{-1/2}-\sigma^{(0)}M/K}{QK\hat{c}_v^{-1/2}}a_u^{(0)}.
\end{align}
These polarization relations are almost identical to atmospheric gravity waves \citep{Schlutow2017b}. 
Only the rescaled Exner pressure variable differs from the polarization of atmospheric gravity waves by the appearance of the extra term $2\mathrm{i}Q^{-1/2}$.
Indisputably, we have found a leading-order difference between the internal centrifugal and the atmospheric gravity waves. 
This result has certainly implications for the interpretation of observations of centrifugal waves by pressure sensors for instance. 
However, for the waves' dynamics and in particular their modulational and stability properties 
it is likely that this difference will play only a minor role.
This claim definitely needs more elaborate investigation which will go beyond the scope of this paper
but will be focussed on in an already envisioned paper.

By the scale assumptions in this regime of intermediate angular velocity 
it is straight-forwardly derived that $u'/C_s=\mathit{O}(\varepsilon)$, 
i.e. the wave-related Mach number is small.
With the aid of this observation and \eqref{eq:rescale_prognostic} 
we can derive the following order relations from the polarization relations
\begin{align}
	\label{eq:scaling_polarization}
	v'/u'=\mathit{O}(1),\quad \theta'/u'=\mathit{O}(\varepsilon),\quad \pi'/u'=\mathit{O}\left(\varepsilon^2\right)
\end{align}
which will become a useful resource in the next section.
%
%\subsubsection*{Anisotropic wave field}

%The leading-order contribution as $\varepsilon\rightarrow 0$
%\begin{align}
%	\hat{c}_v^{-1}Q^2M^2\,\sigma_0^2
%	+\hat{c}_v^{-1}Q^2K^2 = 0 
%\end{align}
%
%\begin{align}
%	\sigma_0^2=\frac{K^2}{M^2}
%\end{align}
%
%\begin{align}
%	\omega^2_\mathrm{hGW}\approx\frac{N^2k^2}{m^2}
%\end{align}

\subsection{Asymptotic regime of high angular frequency}
\label{sec:high}

For the sake of completeness we will study in this section the particular asymptotic regime where $1\gg q=\mathit{O}(\varepsilon)$,
i.e. an extremely fast rotating centrifuge 
where the dimensional scale radii are much smaller than the total radius of the centrifuge.
If we assumed for the moment that $T_0=300$\,K, $r_0=0.5$\,m, $\varepsilon=0.01$ 
and we filled the centrifuge with air, then the rotational frequency would be around 100\,000\,RPM. 

%\subsubsection*{Isotropic wave field}

For this particular regime we define a distinguished limit by
\begin{align}
	q=\varepsilon Q,\quad \kappa=\lceil\varepsilon^{-1}K\rceil,\quad Q,K=\mathit{O}(1)\text{ as } \varepsilon\rightarrow 0
\end{align}
which represents an isotropic scaling. For the anisotropic analysis the reader is referred to the Appendix~\ref{app:high}.
Inserting the distinguished limit into the dimensionless characteristic polynomial \eqref{eq:nondim_charpol}, we obtain to leading order as $\epsilon \rightarrow 0$ four non-trivial roots
\begin{align}
{\sigma^{(0)}}^2 = \frac{K_\mathrm{eff}^2}{2}\left(1\pm\sqrt{1-4\frac{Q^2K^2}{\hat{c}_vK_\mathrm{eff}^4}}\right)
\end{align}
where
\begin{align}
	K_\mathrm{eff}^2=\hat{c}_v^{-1}\left[\hat{c}_p^2/4+Q^2\left(K^2+M^2\right)\right].
\end{align}
We redimensionalize the leading order solution by substituting the scaling assumptions and definitions of the dimensionless variables, so
\begin{align}
\omega^2_\mathrm{AGW} \approx \frac{C_s^2}{2}k_\mathrm{eff}^2\left(1\pm\sqrt{1-4\frac{N_0^2k^2}{C_s^2k_\mathrm{eff}^4}}\right)
\end{align}
where
\begin{align}
	k_\mathrm{eff}^2=k^2+m^2+1/\left(4r_d^2\right),\quad k=\kappa/r_0.
\end{align}
To leading order we obtain the exact same dispersion relation as for atmospheric acoustic-gravity waves (AGW).
Notice that in contrast to the slower regimes the scale separation between the acoustic and the internal mode has vanished.

%\subsubsection*{Anisotropic wave field}

\subsection{Comparison to atmospheric flow regimes}
\label{sec:compare}

To summarize this section, we presented three different asymptotic regimes
that were defined by distinguished limits chosen in terms of the angular frequency of the centrifuge. 
Isotropic wave fields were initially presumed. 
However, the anisotropic scalings were studied as well and are given in the Appendix~\ref{app:aniso}.
%closely relating to atmospheric flow regimes.
We discovered all major atmospheric flow regimes with regard to the strength of stratification 
that are relevant for gravity waves.
Those are the acoustic-gravity waves, scale-separated acoustic and gravity waves encountering amplification, 
Boussinesq gravity waves experiencing no amplification,
as well as the hydrostatic and anelastic versions of these waves.

\begin{figure}
	\centering
	\includegraphics[scale=0.26]{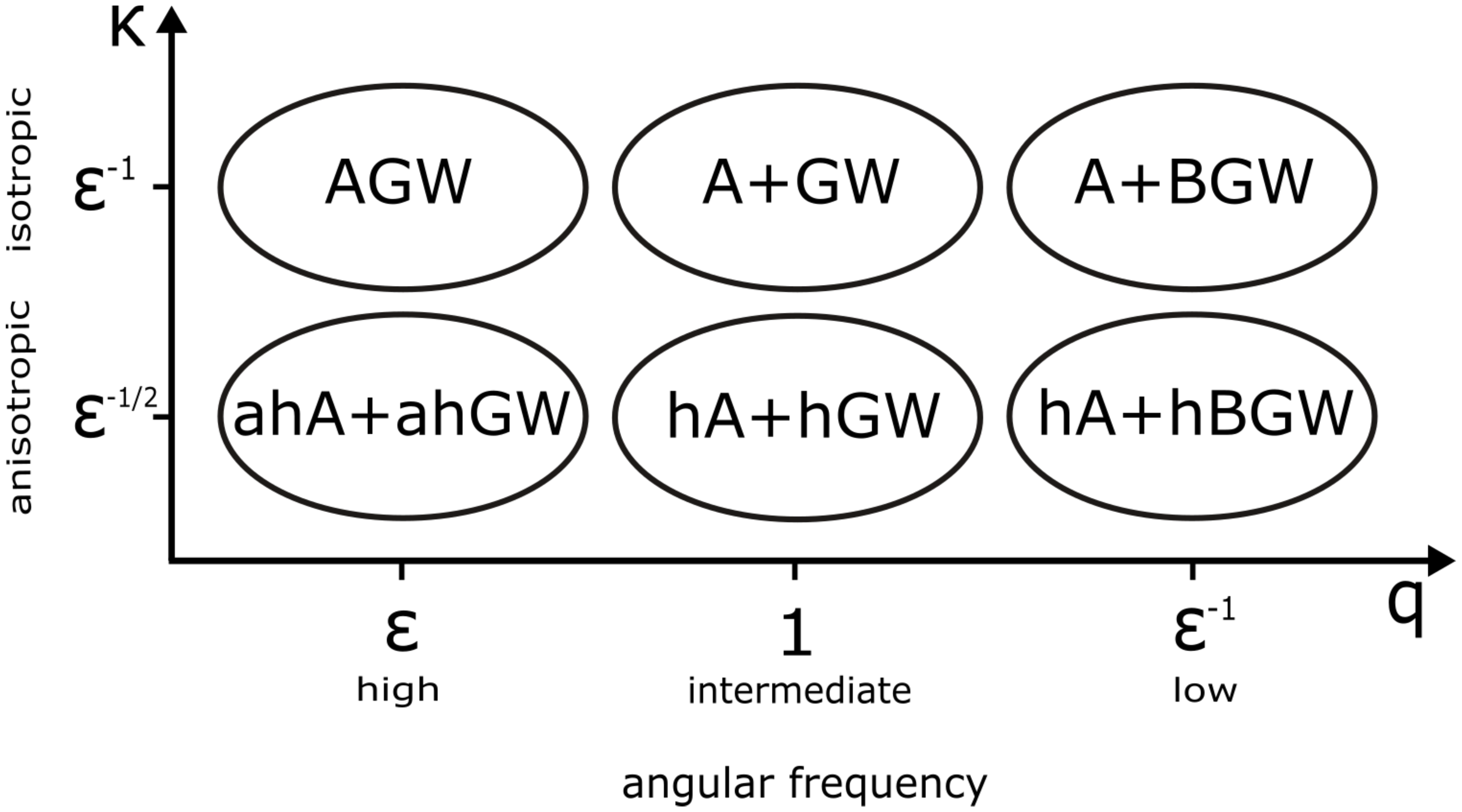}
	\caption{The atmosphere-like asymptotic regimes of the gas centrifuge.
		The abbreviations translate as follows: 
		Acoustic-gravity waves (AGW), 
		acoustic and gravity waves (A+GW), 
		acoustic and Boussinesq gravity waves (A+BGW),
		anelastic hydrostatic acoustic and anelastic hydrostatic gravity waves (ahA+ahGW),
		hydrostatic acoustic and hydrostatic gravity waves (hA+hGW), 
		hydrostatic acoustic and hydrostatic Boussinesq gravity waves (hA+hBGW).
		A plus sign indicates scale separation between two modes.
	}
	\label{fig:regimes}
\end{figure}

In Figure~\ref{fig:regimes}, a diagram giving an overview of the three regimes for both the isotropic as well as anisotropic scaling in terms of the non-dimensional scale radius $q$ and the azimuthal wavenumber $\kappa$ is presented.

\section{Deep waves for the regime of intermediate angular frequency, WKB theory}
\label{sec:wkb}

In the previous section several asymptotic regimes were investigated, all under the assumptions of the shallow-fluid approximation and homogeneity of the wave fields in the axial coordinate by neglect of the gravitational acceleration.
So, the wave dynamics were restricted to a thin fluid layer along the rim of the centrifuge.
We want to relax the assumptions and consider deep three-dimensional waves 
exposed to gravity in the following section 
but restrict ourselves to one particular scaling regime.
From a perspective of atmospheric sciences, the most intriguing regime is the one of intermediate angular frequency. We have argued that this regime closely resembles atmospheric conditions favorable for gravity waves.
The advantage of the shallow-fluid approximation was that the resulting system had constant coefficients allowing for a simple plane wave ansatz. 
For a generalized theory with varying coefficients we want to employ the Wentzel-Kramers-Billouin (WKB) perturbation method. A first step towards its application is the non-dimensionalization of the Euler equations \eqref{eq:govern}.
We choose reference length and time scale for internal waves
based on our findings of the Section~\ref{sec:intermediate},
\begin{align}
	r=\lambda_0\,\hat{r},\quad z=\lambda_0\,\hat{z},\quad 
	t=N_0^{-1}\,\hat{t},\quad N_0=\frac{r_0\Omega^2}{\sqrt{c_pT_0}}
\end{align} 
where $\lambda_0$ denotes the typical wavelength. 
The prognostic variables of the Euler equations are non-dimensionalized by their reference values
labeled by the subscript 0,
\begin{align}
	u=u_0\,\hat{u},\quad v=v_0\,\hat{v},\quad w=w_0\,\hat{w},\quad \theta=\theta_0\,\hat{\theta},\quad \pi=\pi_0\,\hat{\pi}.
\end{align}
For the reference values of the velocities we assume an isotropic wave field such that
\begin{align}
	u_0=v_0=w_0=\lambda_0N_0.
\end{align}
The thermodynamic reference values are set to
\begin{align}
	\theta_0=T_0,\quad\pi_0=(p_0/p_0)^{1/\hat{c}_p}=1.
\end{align}
In contrast to the previous section where we defined a scale separation parameter by the thickness of the shallow layer, we require here that the radial wavelength is small in comparison to the scale radius such that
\begin{align}
	\varepsilon=\frac{\lambda_0}{r_\theta}\ll 1,\quad r_\theta=\frac{c_pT_0}{r_0\Omega^2}.
\end{align}

We recall from Section~\ref{sec:intermediate} that the isotropic regime of intermediate angular frequency is the most interesting regime 
as it has the most resemblance with the atmosphere when it comes to gravity wave dynamics.
This regime allows for radial amplification on the relevant scales and exhibits 
the exact same dispersion relation to leading order 
as atmospheric non-hydrostatic gravity waves. 
Moreover, it shows a clear scale separation to acoustic waves 
and the Coriolis acceleration is negligible.
In line with these arguments, we choose for the non-dimensional scale radius
\begin{align}
	\frac{r_\theta}{r_0}=q=Q=\mathit{O}(1)\text{ as }\varepsilon\rightarrow 0.
\end{align}
Furthermore, we need to determine the strength of the gravitational pull in terms of $\varepsilon$.
For a centrifuge spinning at around 10\,000\,RPM with 50\,cm radius, the gravitational acceleration is approximately five orders of magnitude smaller than the centrifugal acceleration.
If we assume $\varepsilon\approx 0.01$, then a sensible choice for the limit behavior is
\begin{align}
	\frac{g}{r_0\Omega^2}=\mathit{O}(\varepsilon^2)\text{ as }\varepsilon\rightarrow 0.
\end{align}
%Notice that this choice might be an overestimation of the actual strength of gravity as in a real laboratory device $\varepsilon$ is probably chosen to be bigger to obtain larger wavelengths.

By means of the scaling assumptions, the non-dimensional Euler equations become
\begin{subequations}
\label{eq:nondim_govern}
\begin{align}
	\label{eq:nondim_govern_u}
	\varepsilon^2\frac{\partial \hat u}{\partial \hat t} 
	+\varepsilon^2\hat u\frac{\partial \hat u}{\partial \hat r} 
	+\varepsilon^2\frac{\hat v}{\hat r}\frac{\partial \hat u}{\partial \varphi} 
	-\varepsilon^2\frac{\hat{v}^2}{\hat r}
	+\varepsilon^2\hat{w}\frac{\partial\hat{u}}{\partial\hat{z}}
	+Q^2\hat \theta\frac{\partial\hat  \pi}{\partial\hat  r} 
	&= 2\varepsilon^2Q^{1/2}\hat  v + \varepsilon^2Q\hat r \\
	\label{eq:nondim_govern_v}
	\varepsilon^2\frac{\partial \hat v}{\partial \hat t} 
	+\varepsilon^2 \hat u \frac{\partial \hat v}{\partial \hat r} 
	+\varepsilon^2\frac{\hat v}{\hat r}\frac{\partial \hat v}{\partial \varphi} 
	+\varepsilon^2\frac{\hat v \hat u}{\hat r} 
	+\varepsilon^2\hat{w}\frac{\partial\hat{v}}{\partial\hat{z}}
	+Q^2 \frac{\hat \theta}{\hat r}\frac{\partial \hat \pi}{\partial \varphi} &= 
	-2\varepsilon^2 Q^{1/2}\hat u \\
	\label{eq:nondim_govern_w}
	\varepsilon^2\frac{\partial \hat w}{\partial \hat t} 
	+\varepsilon^2\hat u\frac{\partial \hat w}{\partial \hat r} 
	+\varepsilon^2\frac{\hat v}{\hat r}\frac{\partial\hat w}{\partial\varphi}
	+\varepsilon^2\hat{w}\frac{\partial\hat{w}}{\partial\hat{z}}
	+Q^2\hat \theta\frac{\partial\hat\pi}{\partial\hat z}&=-\varepsilon^3Q\\
	\label{eq:nondim_govern_theta}
	\frac{\partial \hat \theta }{\partial \hat t} 
	+\hat u\frac{\partial \hat \theta}{\partial \hat r} 
	+\frac{\hat v}{\hat r}\frac{\partial\hat \theta}{\partial\varphi}
	+\hat{w}\frac{\partial\hat{\theta}}{\partial\hat{z}} &= 0 \\
	\label{eq:nondim_govern_pi}
	\frac{\partial \hat \pi}{\partial \hat t} 
	+\hat u \frac{\partial \hat \pi}{\partial \hat r} 
	+\frac{\hat v}{\hat r} \frac{\partial \hat \pi}{\partial \varphi} 
	+\hat{w}\frac{\partial\hat{\pi}}{\partial\hat{z}}
	+\hat{c}_v^{-1}\hat \pi \left(\frac{1}{\hat r}\frac{\partial(\hat r\hat u)}{\partial \hat r}
	+\frac{1}{\hat r}\frac{\partial \hat v}{\partial \varphi}
	+\frac{\partial\hat{w}}{\partial\hat{z}}\right) &= 0.
\end{align}
\end{subequations}
A stationary, time-independent solution to the non-dimensional governing equations, that will serve as the background for waves, is given by the rigid body rotation of the fluid 
\begin{align}
%	\bigl(\hat u,\,\hat v,\,\hat \theta,\,\hat \pi\bigr)(\hat r,\varphi,\hat t;\varepsilon)=\bigl(0,\,0,\,\hat \Theta,\,\hat \Pi\bigr)(r;\varepsilon)
	\begin{pmatrix}
		\hat u\\ \hat v\\ \hat w\\ \hat \theta\\ \hat \pi
	\end{pmatrix} (\hat r,\varphi,\hat z,\hat t;\varepsilon)
	= \begin{pmatrix}
		0\\ 0\\ 0\\ \hat \Theta\\ \hat \Pi
	\end{pmatrix}(\hat r,\hat z;\varepsilon).
\end{align}
The background variables must solve the radial and axial momentum equations \eqref{eq:nondim_govern_u} and \eqref{eq:nondim_govern_w}, respectively,
\begin{subequations}
\begin{align}
	\label{eq:stat_u}
	Q\hat \Theta\,\frac{d\hat \Pi}{d\hat r} &= \varepsilon^2\,\hat r\\
	\label{eq:stat_w}
	Q\hat \Theta\frac{\partial\hat\Pi}{\partial\hat z}&=-\varepsilon^3.	
\end{align}
\end{subequations}
The radial momentum equation \eqref{eq:stat_u} is closely related to the hydrostatic balance
of the axial momentum equation \eqref{eq:stat_w}
but with the difference of a metric coefficient on the right hand side 
that reflects the curvature of the geometry. 
For the sake of simplicity, we restrict our derivations to an isothermal background, 
so ${\hat \Pi\,\hat \Theta=1}$.
The momentum equations admit an analytical solution under this assumption,
\begin{align}
	\label{eq:nondim_background}
	\hat \Pi=\hat \Pi_c\,\exp\left(\frac{\varepsilon^2\hat{r}^2}{2Q}-\frac{\varepsilon^3\hat z}{Q}\right),\quad
	\hat \Theta=\hat \Theta_c\,\exp\left(-\frac{\varepsilon^2\hat{r}^2}{2Q}+\frac{\varepsilon^3\hat z}{Q}\right).
\end{align}
The coefficients with subscript $c$ represent constants of integration 
which can be fixed by boundary conditions.
It can be observed that the background is only slowly varying on the chosen scales, i.e.
$\hat\Pi(\hat{r},\hat{z};\varepsilon)=\hat\Pi(\varepsilon\hat{r},\varepsilon^3\hat{z})$ and $\hat\Theta(\hat{r},\hat{z};\varepsilon)=\hat\Theta(\varepsilon\hat{r},\varepsilon^3\hat{z})$.
However, the variation in $\hat{r}$ is much stronger than the variation in $\hat{z}$.
Endorsed by these observations, we may use WKB theory 
and introduce compressed coordinates, 
so
\begin{align}
	\rho=\varepsilon\hat{r},\quad
	\zeta=\varepsilon\hat{z},\quad
	\tau=\varepsilon\hat{t}.
\end{align}
The axial re-scaling is chosen accordingly to the dominant radial scale
since the wave field is assumed to be isotropic.
We apply the ansatz
\begin{subequations}
\begin{align}
	\hat{u}&=u'\\
	\hat{v}&=v'\\
	\hat{w}&=w'\\	
	\hat{\theta}&=\hat{\Theta}+\varepsilon\,\theta'\\
	\hat{\pi}&=\hat{\Pi}+\varepsilon^2\,\pi'
\end{align}
\end{subequations}
for the perturbations of the background flow.
%
%\textit{But where is the analastic amplification here? It is buried in $\zeta$ dependency of the $b$s.}

The primed perturbation variables are weighted in terms of powers of $\varepsilon$ according to the order relations \eqref{eq:scaling_polarization}.
Presuming that the perturbations are infinitesimally small,
we linearize the dimensionless Euler equations \eqref{eq:nondim_govern} 
around the background state. 
We make the WKB ansatz for the perturbation taking the slowly varying background into account 
and assuming a comparatively rapid variation of the wave field;
and hence
\begin{align}
	\begin{pmatrix}
		u'\\ v'\\ w'\\ \theta'\\ \pi'
	\end{pmatrix}(\hat{r},\varphi,\hat z,\hat{t};\varepsilon)
	=\begin{pmatrix}
		b_u\\ b_v\\  b_w\\b_\theta\\ b_\pi
	\end{pmatrix}(\rho,\zeta,\tau;\varepsilon)\,
	\exp\left(\mathrm{i}\kappa_\varepsilon\varphi\right)\,
	\exp\left[\mathrm{i}\frac{\phi(\rho,\zeta,\tau)}{\varepsilon}\right],~
	\kappa_\varepsilon=\left\lceil\frac{K}{\varepsilon}\right\rceil
\end{align}
where the azimuthal wavenumber $\kappa_\varepsilon$ is a large number, as $K=\mathit{O}(1)$,
but an integer to ensure periodic solutions in the azimuthal direction.
We introduce the phase function $\phi$ 
and define the slowly varying radial wavenumber and frequency, respectively, via 
\begin{align}
	M(\rho,\zeta,\tau)=\frac{d\phi}{d\rho},\quad
	L(\rho,\zeta,\tau)=\frac{d\phi}{d\zeta},\quad
	\sigma(\rho,\zeta,\tau)=-\frac{d\phi}{d\tau}.
\end{align}
The amplitudes are expanded in a series
\begin{align}
	\boldsymbol{b}(\rho,\zeta,\tau;\varepsilon)=\boldsymbol{b}^{(0)}(\rho,\zeta,\tau)
	+\varepsilon\,\boldsymbol{b}^{(1)}(\rho,\zeta,\tau)
	+\mathit{O}\left(\varepsilon^2\right).
\end{align}
Inserting the assumptions and definitions into the linearized dimensionless Euler equations and collecting terms in powers of $\varepsilon$, we obtain to leading order a linear system of equations that we write in matrix notation as
\begin{align}
	\label{eq:lead_WKB}
	\mathsfbi{N}\,
	\begin{pmatrix}
		b_u^{(0)}\\
		b_v^{(0)}\\
		b_w^{(0)}\\
		Q / \hat{\Theta}\,b_\theta^{(0)}\\
		Q^2\hat{\Theta}\,b_\pi^{(0)}
	\end{pmatrix}=0,\quad
	\mathsfbi{N}=
	\begin{pmatrix}
		-\mathrm{i}\sigma& -2Q^{1/2}& 0& \rho& \mathrm{i}M\\
		2Q^{1/2}& -\mathrm{i}\sigma& 0& 0& \mathrm{i}K/\rho\\
		0& 0& -\mathrm{i}\sigma& 0& \mathrm{i}L\\
		-\rho& 0& 0& -\mathrm{i}\sigma& 0\\
		\mathrm{i}M& \mathrm{i}K/\rho& \mathrm{i}L& 0& 0
	\end{pmatrix}.
\end{align}
To next order we obtain
\begin{align}
	\label{eq:next_WKB}
	\mathsfbi{N}\,
	\begin{pmatrix}
		b_u^{(1)}\\
		b_v^{(1)}\\
		b_w^{(1)}\\
		Q / \hat{\Theta}\,b_\theta^{(1)}\\
		Q^2\hat{\Theta}\,b_\pi^{(1)}
	\end{pmatrix}+
	\begin{pmatrix}
		\frac{\partial}{\partial\tau}& 0& 0& 0& \frac{\rho}{Q}+\frac{\partial}{\partial\rho}\\
		0& \frac{\partial}{\partial\tau}& 0& 0& 0\\
		0& 0& \frac{\partial}{\partial\tau}& 0& \frac{\partial}{\partial\zeta}\\
		0& 0& 0& \frac{\partial}{\partial\tau}& 0\\
		\hat{c}_v\frac{\rho}{Q}+\frac{\partial}{\partial\rho}& 0& \frac{\partial}{\partial\zeta}& 0& 0
	\end{pmatrix}
	\begin{pmatrix}
		b_u^{(0)}\\
		b_v^{(0)}\\
		b_w^{(0)}\\
		Q / \hat{\Theta}\,b_\theta^{(0)}\\
		Q^2\hat{\Theta}\,b_\pi^{(0)}
	\end{pmatrix}
	=0.
\end{align}

%where
%\begin{align}
%	\hat N=\sqrt{-\frac{\rho}{Q\hat{\Theta}}\frac{d\hat{\Theta}}{d\rho}}
%\end{align}
%denotes the dimensionless Brunt-Väisälä frequency.
The leading-order system \eqref{eq:lead_WKB} has a solution 
if and only if the coefficient matrix $\mathsfbi{N}$ is singular
which results in three distinct branches of the dispersion relation
\begin{subequations}
\begin{align}
	\sigma&=0,\\
	\sigma^2&=\frac{\rho^2(K^2/\rho^2+L^2)+4QL^2}{K^2/\rho^2+L^2+M^2}.
\end{align}
\end{subequations}

The next-order system \eqref{eq:next_WKB} is multiplied from the left with the conjugate transposed solution vector of \eqref{eq:lead_WKB}. Since $\mathsfbi{N}$ is skew-Hermitian, the terms acting on $\boldsymbol{b}^{(1)}$ vanish. Taking the real part of the emerging equation yields a prognostic equation,
\begin{align}
	\frac{\partial\hat{E}}{\partial\tau}
	+\frac{\partial}{\partial\rho}\Real\left(\hat{D}\,{b_u^{(0)}}^\ast\,Q^2\hat{\Theta}\,b_\pi^{(0)}\right)
	+\frac{\partial}{\partial\zeta}\Real\left(\hat{D}\,{b_w^{(0)}}^\ast\,Q^2\hat{\Theta}\,b_\pi^{(0)}\right)=0
\end{align}
for the wave energy density 
\begin{align}
	\hat{E}=\frac{1}{2}\hat D\left(\left|b_u^{(0)}\right|^2+\left|b_v^{(0)}\right|^2
	+\left|b_w^{(0)}\right|^2+\left|\frac{Q}{\hat{\Theta}}\,b_\theta^{(0)}\right|^2\right)
\end{align}
%being the sum of kinetic and potential energy 
where
\begin{align}
	\label{eq:nondim_dens}
	\hat{D}(\rho)=\hat{D}_c\exp\left(\frac{\hat{c}_p\rho^2}{2Q}\right)
\end{align}
denotes the non-dimensional background density.
Here, $\Real$ is the symbol for the real part.
The prognostic equation for the wave energy density depicts a conservation law 
as its volume integral is constant in time.
As the energy of a wave packet emitted from the outer rim is conserved, 
the leading-order amplitude of the perturbation must therefore increase when the packet propagates into the interior according to
\begin{align}
	\left(b_u^{(0)},\,
	b_v^{(0)},\,
	b_w^{(0)},\,
	\frac{Q}{\hat{\Theta}}\,b_\theta^{(0)},\,
	Q^2\hat{\Theta}\,b_\pi^{(0)}\right)
	\propto\hat{D}^{-1/2}.
\end{align}
The reasoning is that the energy density composed from the amplitude squared 
is proportional to the background density. 
And thus, the amplitudes must increase with the inverse of the square root of the background density
as all amplitudes are connected by the polarization relation.
In conclusion, we rediscovered the mechanism of radial amplification with the WKB theory, 
which we have already found in the simplified shallow-fluid model.
It is, furthermore, worth noting that we have not found an effect due to gravity in the leading-order results.

Before we summarize the findings of the WKB theory, 
we redimensionalize our results for the dispersion relations yielding
\begin{subequations}
\begin{align}
	\label{eq:triv_bra}
	\omega_r&\approx 0,\\
	\label{eq:gw_bra}
	\omega_r^2&\approx\frac{N_r^2(k_r^2+l^2)+4\Omega^2l^2}{k_r^2+l^2+m^2}
\end{align}
\end{subequations}
where 
\begin{align}
		N_r=\frac{r}{r_\theta}N_0,\quad 
		k_r=\frac{\kappa}{r},\quad 
		l=\frac{L}{\lambda_0},\quad 
		m=\frac{M}{\lambda_0}.
\end{align}
The subscript $r$ reflects the dependency on the radial coordinate.

Let us elaborate on the branches of the dispersion relation.
The trivial branch \eqref{eq:triv_bra} corresponds to a stationary wave solution 
that is referred to as entropy wave in the literature \citep[][and references therein]{Bogovalov2020}.
The nontrivial branches \eqref{eq:gw_bra} look almost exactly 
like the dispersion of inertia-gravity waves in the atmosphere \citep{Achatz2017}. 
The significant difference comes from the inertial term $4\Omega^2l^2$ which depends, 
here, on the axial wavenumber. 
In order to be comparable to the atmosphere it would need to depend on our radial wavenumber $m$ instead.

However, if $l=0$, then \eqref{eq:gw_bra} becomes
\begin{align}
	\omega_r^2&\approx\frac{N_r^2k_r^2}{k_r^2+m^2}
\end{align}
which is the exact same dispersion relation as for atmospheric 
non-hydrostatic gravity waves but in curved geometry due to the metric coefficients.
Let us first note that the dimensional azimuthal wavenumber decreases 
and hence the azimuthal wavelength increases
along the radial axis. 
And second, the dimensional azimuthal phase speed and also the group speed 
as defined by \eqref{eq:phase_group_velo} also grow with increasing $r$
as expected in cylindrical geometry.
We also redimensionalize the solutions \eqref{eq:nondim_background} 
and \eqref{eq:nondim_dens} for the background variables applying the boundary conditions $\Pi(r_0)=\Pi_0,~\Theta(r_0)=\Theta_0$ and $D(r_0)=D_0$, so
\begin{subequations}
\begin{align}
	\Pi(r)&\approx\Pi_0\,\exp\left[\frac{\Omega^2 }{2c_pT_0}\left(r^2-r_0^2\right)\right]\\
	\Theta(r)&\approx\Theta_0\,\exp\left[-\frac{\Omega^2}{2c_pT_0}\left(r^2-r_0^2\right)\right]\\
	D(r)&\approx D_0\,\exp\left[\frac{\Omega^2}{2RT_0}\left(r^2-r_0^2\right)\right].
\end{align}
\end{subequations}
As the variation in the axial coordinate is two orders weaker than in the radial direction,
the background is essentially constant in $z$.
We can deduce that the background is equivalently warped in the radial coordinate 
due to the curved geometry when compared to the atmosphere.
In the latter the background variables depend exponentially on the altitude
whereas in the centrifuge they depend exponentially on the radial coordinate squared.
In conclusion, the internal wave dynamics in the curved centrifuge is isomorphic
to the Eucledian description of atmospheric gravity waves 
with regard to dispersion and therefore phase and group velocity 
at least to leading order and when the flow is axially homogeneous.
Moreover, the waves encounter radial amplification in the centrifuge 
similar to the altitudinal amplification in the atmosphere.  
This phenomenon is not present in the Boussinesq equations.
In fact, our results are consistent with the pseudo-incompressible equations \citep[cf.][]{Achatz2010}.
Starting our derivations from them would have led to the exact same leading-order results.

\section{Conclusion}
\label{sec:conclusion}
\subsection{Summary}

% Full recap, past tense
We have investigated waves as perturbations to a stably stratified gas in a centrifuge
where the stratification resulted from the centrifugal force. 
Under the shallow-fluid approximation,
three times two scaling regimes were studied in terms of perturbation theory. 
The regimes were characterized by their angular frequencies and the proportion of azimuthal to radial wavenumber
such that we were concerned with the regimes of low, intermediate as well as high angular frequency
and isotropic as well as anisotropic wave fields, respectively.
In all six regimes dispersion relations to leading order were derived that closely resemble waves in the 
hydrostatic, stably stratified atmosphere. Those were acoustic waves, gravity waves and their mixed form, acoustic-gravity waves.

One particular regime was identified to be of special interest for the atmospheric sciences as it features---despite the same dispersion---additional characteristics of gravity waves. 
In the regime of intermediate angular frequency, a clear scale separation between acoustic and gravity waves can be observed, the polarization is similar and an influence by the anomalous Coriolis force only appeared as a higher order correction to the dispersion relation. 

% WKB isomorph
The regime of intermediate angular frequency was also studied in more detail by weakening the 
shallow-fluid approximation, taking gravity into account and allowing additionally axial wave propagation. 
By means of WKB theory, we showed that axially homogeneous 
wave fields in cylindrical geometry supported on a domain 
from the center to the rim are isomorphic to non-hydrostatic atmospheric gravity waves in Euclidean geometry.

But most convincing is the fact that waves in the intermediate regime 
encounter radial amplification similar to altitudinal amplification in the atmosphere
which we argued provides an unprecedented opportunity to study 
stability and nonlinear dynamics, in general, of atmospheric compressible waves in a laboratory.

\subsection{Choice of working gas}

\begin{figure}
	\centering
	\includegraphics[scale=0.4, angle=270]{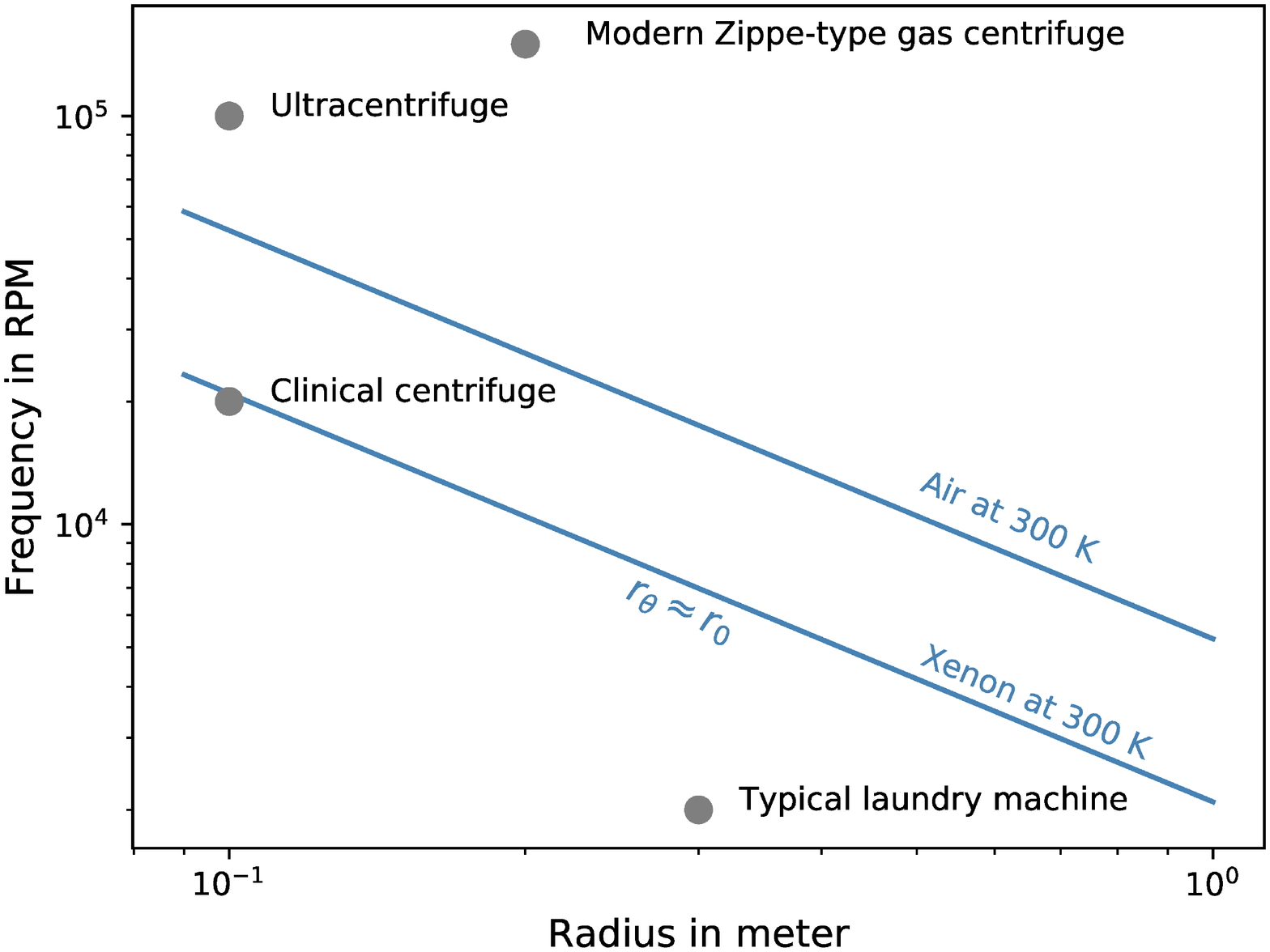}
	\caption{Diagram of radius and rotational frequency. 
		The blue lines correspond to the regime of intermediate angular frequency 
		where the non-dimensional scale radius is ${q=\mathit{O}(1)}$ for air and xenon, respectively. 
		The gray dots represent commercially available centrifuges for the sake of comparison.}
	\label{fig:comparison}
\end{figure}
In order to provide additional evidence for the practicability of the experiment, we illustrate the regime of intermediate angular frequency in Figure~\ref{fig:comparison}. Given a total radius of the centrifuge~$r_0$, the necessary rotational frequency $f$ to achieve this regime is plotted for two different gases, air and xenon. 
The curves are given by setting the non-dimensional scale radius to unity corresponding to the assumption 
that one potential temperature scale radius fits into the centrifuge. 
Then, the rotational frequency becomes ${f(r_0)=\sqrt{c_pT_0}/(2\pi r_0)}$.
The background temperature is ${T_0=300}$\,K.
% best gas for experiments
It is the obvious and most convenient choice to use air as a working gas having ${c_p=1006\,\mathrm{J}\,\mathrm{kg}^{-1}\,\mathrm{K}^{-1}}$.
Since high rotational frequencies generate a lot of stress onto the material, 
lower frequencies for an experiment would be desirable. 
A lower frequency might be achieved by cooling the experiment or alternatively using a gas of low heat capacity
such as xenon which exhibits a heat capacity at constant pressure as low as ${c_p=160\,\mathrm{J}\,\mathrm{kg}^{-1}\,\mathrm{K}^{-1}}$.
By using xenon the frequency is more than halved.

\subsection{Role of dissipation}

The motivation for our investigation was, first and foremost, 
to study the nonlinear dynamics of atmospheric gravity waves and in particular instability processes in a centrifuge. 
For this purpose we studied the compressible Euler equations. 
The question remains whether dissipation is really negligible, at least to leading order.
If any instability growth rates due to the nonlinear dynamics were smaller than the dissipative damping,
then a laboratory centrifuge to study the compressible waves would be rendered unfeasible.
To answer this question we consider the Reynolds number, $\mathit{Re}=\rho_0u_0\lambda_0/\mu$ 
where $\mu$ denotes the dynamic viscosity. 
By means of the scaling assumptions of the previous section, we can estimate for the regime of intermediate angular frequency that
\begin{align}
	\mathit{Re}=\mathit{O}\left(\varepsilon^2\frac{p_0}{\mu\Omega}\right).
\end{align}
Let us set the pressure at the outer wall to atmospheric pressure, $p_0=10^5$\,Pa.
The dynamic viscosity for air or xenon is approximately $\mu\approx 2\cdot 10^{-5}$\,Pa\,s.
The angular frequency in the intermediate regime for a centrifuge with radius $\approx 50$\,cm is $\Omega=2\pi f$ with $f\approx 10\,000$\,RPM.
Consequently, we obtain
\begin{align}
	\frac{p_0}{\mu\Omega}\approx 10^7.
\end{align}
For instance choosing $\varepsilon\approx 0.01\dots 0.1$, we get therefore $\mathit{Re}\approx 10^3\dots 10^5$.
The Reynolds number is large.
We can conclude from these estimates that only for extremely small $\varepsilon$, 
i.e. very small wavelengths in comparison to the scale height, the Reynolds number becomes small.
And therefore viscous effects are indeed negligible. However, they may appear as higher-order corrections 
which needs to be addressed in future investigations.

\subsection{Excitation of waves in the laboratory centrifuge}

An important issue regarding the feasibility of a laboratory gas centrifuge to study atmospheric gravity waves is certainly the excitation mechanism. How do we want to excite waves in a controlled repeatable fashion?
We want to propose two strategies both having their own benefits and drawbacks.
\begin{enumerate}
	\item The basic idea is to add an axially homogeneous orography undulating with the desired azimuthal wavenumber to the outer rim. In order to excite a wave with defined properties, a steady azimuthal flow over the orography could be generated like the horizontal wind over a mountain range in the real atmosphere that results in lee waves. 
	Exploiting the inertia of the gas in rigid body rotation, we may induce such a flow by bringing the centrifuge slowly to a certain angular frequency assuring the rigid body rotation. 
	And then, by a sudden acceleration to the target angular frequency, inertia creates the envisioned flow over the orography and excites waves.
	This strategy demands a strong and precisely controllable motor driving the centrifuge.
	Deceleration is certainly easier to achieve as a brake needs to be readily applied.
	However, deceleration of the rigid body flow may give rise to centrifugal instability \citep{Kundu2002} which we want to avoid.
	\item Another idea also relies on the undulating orography at the outer rim. But instead of generating a differential velocity by inertia, it is alternatively possible to induce a background shear flow by letting the outer cylinder rotate with a different angular frequency from the inner cylinder.
	This strategy is similar to the configuration of the Taylor–Couette flow which has frequently been realized with high rotational velocities.
	On the one hand, this type of configuration may be easier to control and even more realistic
	as usually the flow over mountains is also vertically sheared.
	On the other hand, the wave solutions for a radially sheared background flow are more complicated than what was presented in the theoretical part of this study.
	However, if the shear is not too strong, then WKB theory remains equally applicable
	and wave solutions can be obtained.
\end{enumerate} 

\subsection{Observation and measurement of the wave field in the centrifuge}

Another important issue that needs to be addressed is the measurement of the wave field inside the centrifuge.
The biggest challenge in this regard are the strong centrifugal forces of about 10\,000\,$g$ 
making the survival of measuring probes in the rotating system questionable. 
Hence, ways of observing the wave field from the outside are desirable.
A possibility would be the use of laser light from a source in the stationary frame 
traversing the gas in the centrifuge in the axial direction through inspection windows in the top and bottom lid of the centrifuge. 
From the brightness of the laser beam, density could be inferred. 
The measured density would be an integrated quantity over the entire length of the centrifuge.
Since the wave fields are conceived to be axially homogeneous, 
it would still, however, be feasible to reconstruct the complete flow field from the density
due to the polarization relations.

% 3-dimesionality
%Throughout this paper we have assumed an axially homogeneous flow, so ${\partial/\partial z~\cdot=0}$. 
%The extension towards a fully three-dimensional wave field is straight forward
%as the base state, i.e. the isothermal gas in rigid body rotation, does not depend on the axial coordinate.
% Future plans: sheared flow as background. 
%Note that beyond the rigid body rotation, a flow with radially sheared azimuthal velocity 
%also solves the stationary Euler equations. If the shear is not too strong,
%then WKB theory remains applicable.

\subsection{Final remarks}

% Inclusion of dissipation.
For future studies, one might also add dissipation asymptotically to the picture 
and lift the assumption of small amplitudes to a fully nonlinear wave theory similar to \cite{Achatz2010,Schlutow2017b}.

% Stratified/2d-turbulence. 
%We want to give a final remark on the gas centrifuge as an experimental device.
The gas centrifuge as an experimental device might be also used to study stratified turbulence which is a common theme in the atmospheric sciences.
The role of compressibility for two-dimensional turbulence may be explored for the first time in a laboratory with a repeatable experiment.

	\section*{Acknowledgments}
	This research was supported by the German Research Foundation (DFG) through Grant KL 611/25-2
of the Research Unit FOR 1898.	
	
	\appendix	
	\section{Anisotropic wave fields}
\label{app:aniso}

This appendix continues the Section~\ref{sec:shallow} on waves in the shallow-fluid approximation.
In the main text, we assumed isotropic wave fields: equal radial and azimuthal scale; no favorable direction.
It is very common in the atmosphere that the scales do differ. 
When the horizontal scale is much longer than the vertical scale, 
the vertical momentum equation can be approximated by the hydrostatic equation.
The resulting wave solutions are therefore called hydrostatic waves 
which are a kind of anisotropic waves.

\subsection{Regime of low angular frequency}
\label{app:low}

In this paragraph we revisit the regime of low angular frequency of Section~\ref{sec:low} and define an
distinguished limit by
\begin{align}
	q=\varepsilon^{-1}Q,\quad \kappa=\lceil\varepsilon^{-1/2}K\rceil,\quad Q,K=\mathit{O}(1)\text{ as } \varepsilon\rightarrow 0.
\end{align}
Note that due to the assumptions we got $\kappa/\mu=\mathit{O}(\varepsilon^{1/2})$
and therefore the azimuthal wavenumber is much smaller than the radial wavenumber, 
similar to the hydrostatic scaling in the governing equations for the atmosphere. 
This particular distinguished limit generates an anisotropic wave field
since the azimuthal axis is elongated in comparison to the radial axis.

When we insert our scaling assumptions into the dimensionless characteristic polynomial \eqref{eq:nondim_charpol} 
and pass to the limit $\varepsilon\rightarrow 0$, we only find two trivial roots.
Two complementary roots escape to infinity. Consequentially, we face a singular perturbation problem. 
First, the two trivial roots are recovered by the rescaling 
\begin{align}
	\sigma=\varepsilon^{1/2}\Sigma.
\end{align}
The rescaled characteristic polynomial reads
\begin{align}
	\hat{c}_v\varepsilon^5\Sigma^4-\left[\varepsilon^4\hat{c}_p^2/4
	+4\varepsilon^3\hat{c}_vQ+Q^2\left(\varepsilon K^2+M^2\right)\right]\,\Sigma^2
	-4\varepsilon^{2}\hat{\Gamma}Q^{3/2}K\,\Sigma+Q^2K^2=0
	\label{eq:aniso_bgw_charpol}
\end{align}
which has the two roots
\begin{align}
	{\Sigma^{(0)}}^2=\frac{K^2}{M^2}.
\end{align}
When redimensionalized by substitution of the dimensional variables in combination with the scaling assumptions, we find
\begin{align}
	\omega^2_\mathrm{hBGW}\approx\frac{N_0^2k^2}{m^2}
\end{align}
which is equivalent to the dispersion relation of hydrostatic Boussinesq gravity waves (hBGW).

Second, the two roots at infinity of the original scaling can be found by the rescaling 
\begin{align}
	\sigma=\varepsilon^{-2}\Sigma.
\end{align}
Substituting and passing to the limit results in two non-trivial roots
\begin{align}
	{\Sigma^{(0)}}^2=\hat{c}_v^{-1}Q^2M^2
\end{align}
which yield in their dimensional form
\begin{align}
	\omega^2_\mathrm{hA}\approx C_s^2m^2.
\end{align}
These roots compare to hydrostatic acoustic waves in the atmosphere (hA).

\subsection{Regime of intermediate angular frequency}
\label{app:inter}

In this paragraph we explore the anisotropic wave field in the regime of intermediate angular frequency
from Section~\ref{sec:intermediate}. 
We assume the following distinguished limit
\begin{align}
	q=Q,\quad \kappa=\lceil\varepsilon^{-1/2}K\rceil,\quad Q,K=\mathit{O}(1)\text{ as } \varepsilon\rightarrow 0.
\end{align}
Substituting the distinguished limit in the dimensionless characteristic polynomial \eqref{eq:nondim_charpol} yields a singular perturbation problem
where there are two vanishing roots and two additional roots that tend to infinity.
To recover the vanishing roots, we rescale the frequency ${\sigma=\varepsilon^{1/2}\Sigma}$ which to leading order results in
\begin{align}
	{\Sigma^{(0)}}^2=\frac{K^2}{M^2}.
\end{align}
When we rewrite this equation in terms of the dimensional variables, 
\begin{align}
	\omega^2_\mathrm{hGW}\approx\frac{N_0^2k^2}{m^2},
\end{align}
we obtain the exact same dispersion relation as for atmospheric hydrostatic gravity waves (hGW).
Next, let us seek the roots that escaped to infinity. 
The appropriate rescaling is determined by $\sigma=\varepsilon^{-1}\Sigma$.
Two non-vanishing roots can be found to leading order employing the approach in the characteristic polynomial,
\begin{align}
	{\Sigma^{(0)}}^2=\hat{c}_v^{-1}Q^2M^2.
\end{align}
In its dimensional form the dispersion relation reads
\begin{align}
	\omega_\mathrm{hA}^2\approx C_s^2m^2
\end{align}
which is equivalent to hydrostatic acoustic waves in the atmosphere (hA).

\subsection{Regime of high angular frequency}
\label{app:high}

Concluding the appendix we will investigate the anisotropic wave field for the regime of high angular frequency
from Section~\ref{sec:high}. The following distinguished limit is assumed
\begin{align}
	q=\varepsilon Q,\quad \kappa=\lceil\varepsilon^{-1/2}K\rceil,\quad Q,K=\mathit{O}(1)\text{ as } \varepsilon\rightarrow 0.
\end{align}
We insert the distinguished limit into \eqref{eq:nondim_charpol} 
and obtain to leading order two vanishing and two non-vanishing roots
\begin{align}
	{\sigma^{(0)}}^2=\hat{c}_v^{-1}\left(\hat{c}_p^2/4+Q^2M^2\right)
\end{align}
which reads in dimensional variables
\begin{align}
	\omega_\mathrm{ahA}^2\approx C_s^2\left[m^2+1/\left(4r_d^2\right)\right].
\end{align}
This dispersion relation is equivalent to hydrostatic acoustic waves with an extra term
that represents a correction due to the compressibility of the gas. 
We may refer to this term as anelastic correction (ahA).
The two vanishing roots cannot be assessed by expanding $\sigma$ in terms of $\varepsilon$ 
rendering the perturbation problem to be a singular one.
Therefore, we apply the method of dominant balance that provides the rescaling ${\sigma=\varepsilon^{1/2}\Sigma}$.
To leading order we obtain subsequently
\begin{align}
	{\Sigma^{(0)}}^2=\frac{Q^2K^2}{\hat{c}_p^2/4+Q^2M^2}.
\end{align}
Redimensionalization of the dispersion relation yields
\begin{align}
	\omega_\mathrm{ahGW}^2\approx\frac{N_0^2k^2}{m^2+1/\left(4r_d^2\right)}
\end{align}
which is equivalent to the dispersion relation of anelastic hydrostatic gravity waves (ahGW) in the atmosphere.
The reasoning to call the waves anelastic comes from the fact that the wave solutions of the anelastic equations by \cite{Lipps1982} together with the hydrostatic approximation give the same dispersion relation.
We observe that in contrast to the isotropic regime of high angular frequency a
scale separation between acoustic and internal modes is re-established.
However, the scale separation is only of $\mathit{O}\left(\varepsilon^{1/2}\right)$ 
being comparatively weak.

% 	\clearpage

%	\printbibliography[heading=bibintoc]
	\bibliographystyle{abbrvnat}
	\bibliography{library.bib}

\end{document}